\documentclass[lettersize,journal]{IEEEtran}
\usepackage{amsmath,amsfonts}
\usepackage{algorithmic}
\usepackage{algorithm}
\usepackage{array}
\usepackage[caption=false,font=small,labelfont=rm,textfont=rm]{subfig}
\usepackage{textcomp}
\usepackage{stfloats}
\usepackage{url}
\usepackage{verbatim}
\usepackage{graphicx}
\usepackage{cite}
\usepackage{float}
\usepackage{booktabs}
\usepackage{amsmath}
\usepackage{amssymb}
\usepackage{booktabs} 
\usepackage{graphicx} 
\usepackage{multirow} 
\usepackage{enumitem}
\usepackage{tabularx}  
\usepackage{array} 

\begin{document}

\title{Transfer-based Adversarial Poisoning Attacks \\ for Online (MIMO-)Deep Receviers}

\author{ Kunze Wu, Weiheng Jiang,~\IEEEmembership{Member,~IEEE}, Dusit Niyato,~\IEEEmembership{Fellow,~IEEE}, Yinghuan Li, and Chuang Luo
 }



\maketitle

\begin{abstract}

  Recently, the design of wireless receivers using deep neural networks (DNNs), 
  known as deep receivers, has attracted extensive attention for ensuring reliable communication in complex channel environments. 
  To adapt quickly to dynamic channels, online learning has been adopted to update the weights of deep receivers with over-the-air data (e.g., pilots). 
  However, the fragility of neural models and the openness of wireless channels expose these systems to malicious attacks. 
  To this end, understanding these attack methods is essential for robust receiver design.
  In this paper, we propose a transfer-based adversarial poisoning attack method for online receivers.
  Without knowledge of the attack target, adversarial perturbations are injected to the pilots, poisoning the online deep receiver and impairing its ability to adapt to dynamic channels and nonlinear effects. 
  In particular, our attack method targets Deep Soft Interference Cancellation (DeepSIC)\cite{shlezinger2020deepsic} using online meta-learning.
  As a classical model-driven deep receiver, DeepSIC incorporates wireless domain knowledge into its architecture. 
  This integration allows it to adapt efficiently to time-varying channels with only a small number of pilots, achieving optimal performance in a multi-input and multi-output (MIMO) scenario.
  The deep receiver in this scenario has a number of applications in the field of wireless communication, which motivates our study of the attack methods targeting it.
  Specifically, we demonstrate the effectiveness of our attack in simulations on synthetic linear, synthetic nonlinear, static, and COST 2100 channels. 
  Simulation results indicate that the proposed poisoning attack significantly reduces the performance of online receivers in rapidly changing scenarios.


\end{abstract}


\section{Introduction}

\IEEEPARstart{I}{n} recent years, the application of deep learning (DL) in designing wireless communication systems has garnered significant interest. 
Researchers have concentrated on employing DL in wireless receivers to enhance communication performance in complex channels 
and to bolster adaptability in dynamic environments
\cite{alexandropoulos2023hybrid,shlezinger2019asymptotic}. 
However, DL-based wireless applications face vulnerabilities to evasion and data poisoning attacks 
owing to the inherent openness of wireless channels and the fragility of neural models \cite{adesina2022adversarial}. 
Investigating attack methodologies on deep receivers serves to elucidate their response under such threats, 
thereby facilitating the development of secure wireless DL systems, which forms the primary focus of this paper.

\subsection{DL-based Wireless Receivers and Related Applications}

Till now, numerous studies have explored DL-based designs for wireless communication systems. 
Most of them utilize an independent DNN to map the input-output relationships of functional modules in communication links.
Alternatively, some approaches jointly optimize modules at both the transmitter and receiver by multiple DNNs.
Typical applications include DL-based adaptive modulation\cite{aoudia2022waveform}, channel estimation\cite{fu2023deep}, 
channel coding and decoding\cite{farsad2018deep}, and modulation recognition\cite{liu2020modulation}.
Beyond replacing functional modules in the physical layer, 
various constraints can be integrated into the training process to optimize additional system metrics, 
such as the adjacent channel leakage ratio (ACLR) and peak-to-average power ratio (PAPR) \cite{aoudia2022waveform}.
In terms of receiver design, lots of work has contributed to enhancing its adaptability to the dynamic channels, 
including training multiple models as a deep ensemble \cite{raviv2020data} and joint learning 
\cite{o2017introduction,raviv2024adaptive}.

However, DL-based wireless designs mentioned above are data-driven methods 
that heavily depend on a substantial amount of training data to improve their generalization capacity. 
Given that deep receivers typically have access to only limited pilots for adaptation, this characteristic poses a significant challenge.
In addition, data-driven designs may suffer from performance degradation 
when faced with data distribution drift caused by dynamic channels\cite{park2020learning}.
To handle these issues, an online learning based approach has been proposed, involving dataset, training algorithm and deep receivers architecture.
In particular, data augmentation\cite{raviv2023data} and self supervision methods
\cite{Finish_Cohen_Raviv_Shlezinger_2022},\cite{Shlezinger_Farsad_Eldar_Goldsmith_2020} were proposed to to expand  the training data for online adaptation. 
In \cite{park2020learning} and \cite{raviv2023online}, meta-learning was employed to improve the generalization capability.
Furthermore, \textit{Model-based deep learning} provide a solution for receivers architecture design 
to satisfy the adaptability and data efficiency\cite{shlezinger2023model}.  
Specifically, the deep receivers were explicitly modelled by incorporating wireless domain knowledge, thereby reducing the dependence on data, 
such as DNN-aided inference\cite{raviv2024adaptive},\cite{shlezinger2023model} and deep unfolding
\cite{shlezinger2023model,raviv2024adaptive,balatsoukas2019deep,shlezinger2020deepsic}. 
In these studies, \cite{shlezinger2020deepsic} proposed a classic model-based deep receiver, i.e., the DeepSIC, in MIMO scenarios, 
derived from the iterative soft interference cancellation (SIC)\cite{choi2000iterative} MIMO detection algorithm. 
It employed DNN instead of each round of interference cancellation and soft detection, 
requiring only a few iterations to achieve extremely low data dependence and optimal performance. 
\cite{raviv2023online} utilized meta-learning to improve the training performance of online DeepSIC,
and the evaluation results indicated that its performance was improved compared with traditional data-driven receiver, 
and it exhibited commendable adaptability to dynamic channels.

\subsection{Security of DL in Wireless Communications}

As mentioned earlier, while DL-based transceiver designs can enhance performance, they remain vulnerable to attacks by malicious users.
In particular, attacks on DL-based transceivers are divided into two main categories, i.e., the evasion attacks and data poisoning attacks. 
Evasion attacks, also known as adversarial attacks, manipulate test data to mislead the model\cite{adesina2022adversarial},\cite{demontis2019adversarial}. 
On the other hand, data poisoning attacks corrupt the training data, affecting the model's performance during testing\cite{adesina2022adversarial},
\cite{demontis2019adversarial,munoz2017towards,wang1808data}.

According to extensive literature review, numerous studies on DL-based wireless communication primarily concentrate on evasion attacks.
For instance, \cite{bahramali2021robust} proposed generative adversarial network (GAN)-based method to generate adversarial perturbations for channel received data, 
which can unnoticeably mislead wireless end-to-end autoencoders, 
modulation pattern recognition, and the DL-based symbol detection in orthogonal frequency division multiplexing (OFDM) systems. 
\cite{sofer2023interpretable} reported adversarial perturbations can interfere with gradient-based iterative optimization algorithms in the physical layer.
\cite{nan2023physical} proposed semantic attacks against semantic communication.
Furthermore, adversarial perturbations can also play a role in interpretable (e.g., deep unfolding-based) architectures. 
To illustrate, \cite{wang2020adversarial} employed transfer-based methods to attack deep sparse coding networks 
and demonstrated that these attacks exert deleterious effects on the various components of deep unfolded-based sparse coding.
Regarding data poisoning attacks, current research primarily focuses on cognitive radio spectrum-aware poisoning \cite{adesina2022adversarial},\cite{sagduyu2019adversarial} 
and disrupting distributed wireless federated learning\cite{pang2021accumulative,xia2023poisoning}.

\subsection{Contribution of This Paper}
Unlike previous studies, this paper addresses security threats to online deep receivers. 
Furthermore, we propose a transfer-based adversarial poisoning attack method, 
which can significantly corrupt various online deep receivers even without prior knowledge of the target system.
Specifically,
we focus on online receivers based on model-based deep learning, such as DeepSIC\cite{shlezinger2020deepsic} and Meta-DeepSIC\cite{raviv2023online}, 
as well as general DL-based detectors, including the black-box DNN detector\cite{raviv2024adaptive},\cite{raviv2023data} 
and the ResNet detector designed based on DeepRX\cite{honkala2021deeprx}.

As previously stated, DeepSIC is a classical model-based deep receiver that can be combined with meta-learning for efficient online adaptation.
This design effectively tackles the challenge of limited pilot data in wireless communication scenarios, 
thereby improving the generalization of deep receivers under dynamic channel conditions. 
Moreover, studies on the attack methods for DeepSIC can provide comprehensive insights into deep receiver characteristics and contribute to robust designs. 
Ultimately, this research aids in creating secure and efficient DL-enabled wireless communication systems.

Specifically, the mainly contributions of this paper are summarised as below.

\begin{itemize}
  \item[$\bullet$] 
  We highlight a communication system susceptible to malicious user poisoning attacks. 
  We then analyze the vulnerability of the deep receiver based on online learning in the authorization system. 
  From the perspective of malicious user, we further develop an attack utility model and an optimal attack utility decision problem.

  \item[$\bullet$] 
  We effectively design a poisoning attack framework and attack perturbation generation method for online learning deep receivers. 
  The fundamental concept is to introduce a poisoned sample into the online training and updating phase of the deep receiver, 
  thereby compromising its performance over time. The poisoning attack framework has two stages. 
  Firstly, malicious user employ joint learning to create a surrogate model, 
  which can be selected from a generic DNN architecture, e.g., feedforward DNNs. 
  Secondly, they generate poisoning perturbation samples based on the surrogate model. 
  The transferability of the poisoning attack makes it work on different types of deep receivers.

  \item[$\bullet$]We numerically evaluate the effect of the proposed poisoning attack method on four channel models: 
  Linear synthetic channel, nonlinear synthetic channel, static channel, and COST 2100 channel. 
  Simulation results demonstrate that the proposed poisoning attack method impairs the deep receiver's ability to adapt to rapid changes 
  in dynamic channels and to learn from nonlinear effects. 
  Furthermore, deep receivers adapted using meta-learning more severely damaged after poisoning.
\end{itemize}

The rest of the paper is organized as follows. 
Section II introduces the system and scenario models 
and the attack models of the malicious user.
Section III presents the basic theory of adversarial machine learning, focusing on evasion attacks, data poisoning attacks, 
and the conceptual approaches to attack transferability. 
Section IV details the proposed poisoning attack framework and the method for generating poisoning attack samples for online deep receivers. 
Section V evaluates and analyzes the effectiveness of the proposed poisoning attack method. 
Section VI concludes the paper.

\section{System and scenario modelling}
In this section, we first present the communication system and scenario model under the presence of a malicious poisoning attack user in Section II-A. 
Subsequently, we introduce the operational model of the legitimate receiver based on deep learning in Section II-B. 
Finally, we discuss the detail of malicious user poisoning attack, focusing on pilot poisoning attacks in Section II-C.

\begin{figure}[!h] 
  \centering 
  \includegraphics[width=0.48\textwidth]{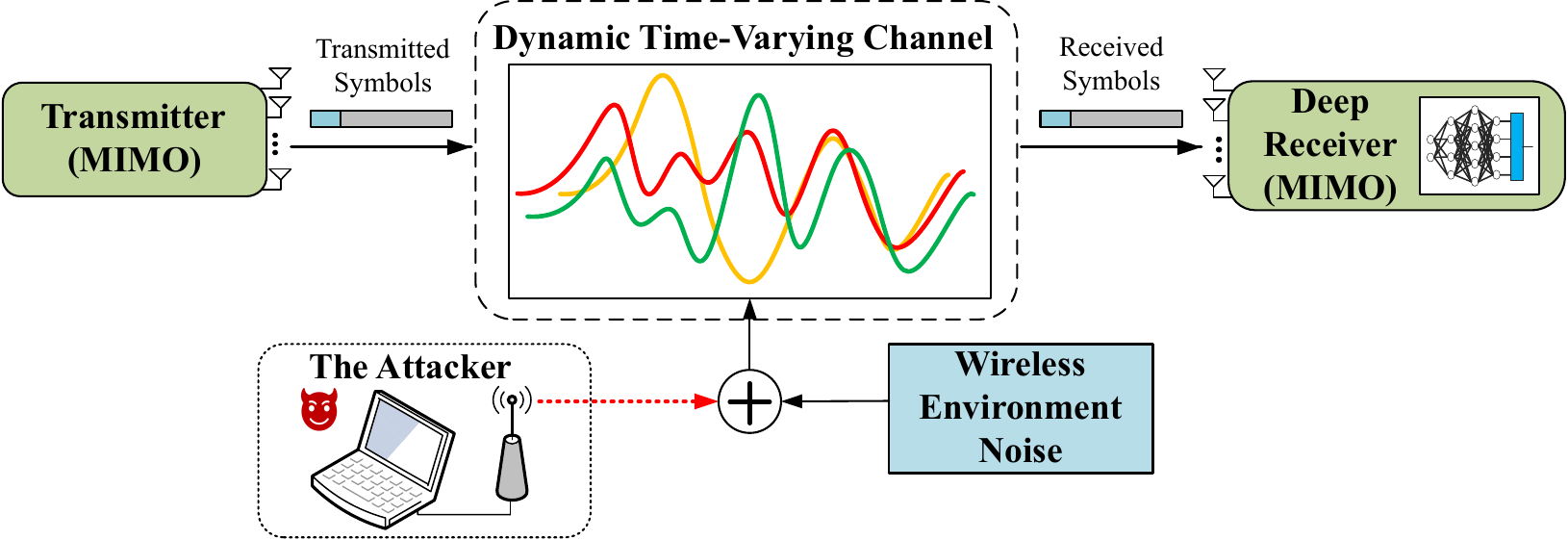} 
  \caption{Scenario modelling of communication systems with malicious user.}
  \label{fig_1} 
\end{figure}

\subsection{Communication System Scenario Model}

In this paper, we investigate a poisoning attack scenario model for a communication system, as illustrated in Fig. 1. 
This system consists of a pair of legitimate transceivers and a malicious poisoning attack user. 
Both the legitimate transmitter and receiver equip with multiple antennas, denoted as $N_{tx}$ and $N_{rx}$ respectively.
We focus on a single-antenna malicious user, as this represents a cost-effective and straightforward approach to conducting attacks.
The data transmission from transmitter to receiver is block-based, as illustrated in Fig. 2. 
The length of one block is $L$, including $L_\text{pilot}$ pilot symbols 
and $L_\text{info}$ information symbols, herein, $L_\text{info} \gg L_\text{pilot}$.
As shown in Fig. 3, the legitimate receiver utilizes a DL-based architecture for signal receiving and processing. 
It is trained using pilot data and employs the trained deep receiver to decode information data.
For the malicious user, based on the previously collected
pilot data, it launches an attack by  poisoning or disturbing the transmission process of the pilot used by the legitimate user. 
Its objective is to corrupt with the online training and updating of the deep receiver, 
thereby disrupting its information data reception and decoding.

\begin{figure}[!t]
  \centering
  \includegraphics[width=0.4\textwidth]{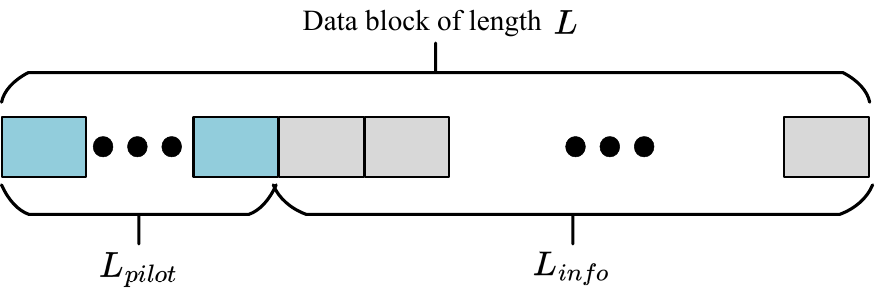}
  \caption{Transmitted data format.}
  \label{fig_2}
\end{figure}

Based on the above illustrated scenario,
defining the transmit symbols by the legitimate transmitter as
$\mathbf{s} \in \mathbb{R}^{N_{\mathrm{tx}}}$, 
and the corresponding modulated symbols as
$\mathbf{x} \in \mathbb{C}^{N_{\mathrm{tx}}}$. 
These modulation symbols are then upconverted, amplified, transmitted through multiple antennas, 
and finally arrived at the receiver.
Let $\mathbf{H} \in \mathbb{R}^{N_{\mathrm{rx}} \times N_{\mathrm{tx}}}$
denote the baseband equivalent channel matrix, and
$\mathbf{w} \sim \mathcal{C N}\left(\mathbf{0}, \sigma^2 \ \mathbf{I}\right)$ 
represent the additive gaussian white noise experienced by the legitimate receiver.
The equivalent baseband signal received by the legitimate receiver, 
in the absence of a malicious user poisoning attack, is
$\mathbf{y} \in \mathbb{C}^{N_{\mathrm{rx}}}$, 
which can be expressed as

\begin{equation} \label{eq:1}
  \mathbf{y}=\mathbf{H x}+\mathbf{w} \text{. }
\end{equation}

Following that, defining a block of data symbols received by the authorised receiver as $\mathbf{Y}$. 
The received data symbols can then be divided into two parts, denoted as
$\mathbf{Y}_{\text{pilot}}=\left\{\mathbf{y}_i\right\}_{i=1}^{L_{\text{pilot}}}$ 
and $\mathbf {Y}_\text{info}=\left\{\mathbf{y}_i\right\}_{i=L_{\text{pilot}}+1}^L$.

\subsection{Deep Receiver with Online Training}

\begin{figure}[!h] 
  \centering 
  \includegraphics[width=0.4\textwidth]{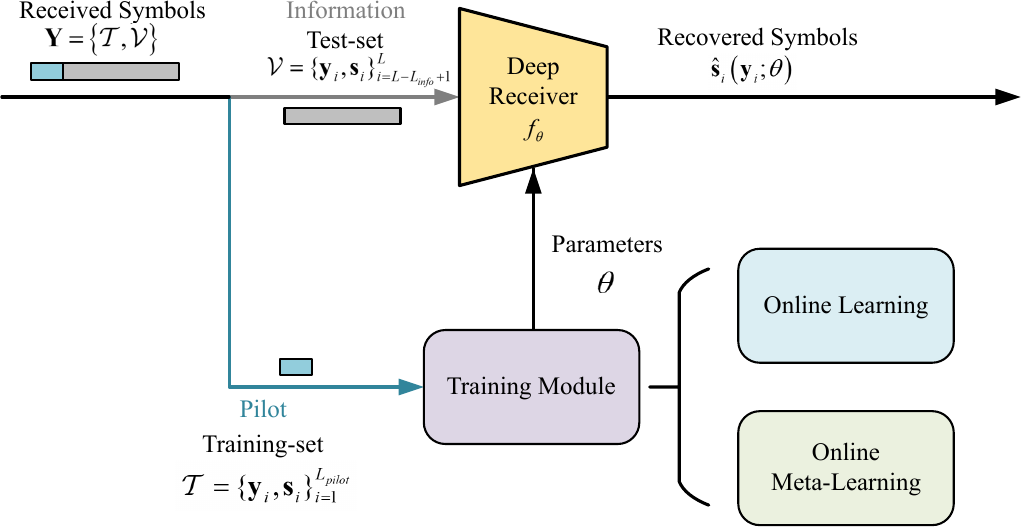} 
  \caption{Online deep receiver workflow.}
  \label{fig_3} 
\end{figure}

In this paper, 
the online learning based deep receiver is adopted
by the legitimate receiver, as illustrated in Fig. 3.
Here, define the deep receiver as a classifier $f_\theta$ 
with model parameter $\theta$. $f_\theta$ is trained using a supervised learning approach.
The data used for training is the pilot dataset, which is defined as 
$\mathcal{T}=\left\{\mathbf{Y}_{\text {pilot}}, \mathbf{S}_{\text {pilot}}\right\}=\left\{\mathbf{y}_i, \mathbf{s}_i\right\}_{i=1}^{L_{\text {pilot}}}$. 
Model testing is done with information dataset, which is represented as 
$\mathcal{V}=\left\{\mathbf{Y}_{\text {info}}, \mathbf{S}_{\text {info}}\right\}=\left\{\mathbf{y}_i, \mathbf{s}_i\right\}_{i=L_{\text {pilot}}+1}^L$. 
The supervised training loss function is the cross-entropy loss, which is represented as  
$\mathcal{L}(\mathcal{T} ; \theta)$.  
$\hat{P}_\theta(\cdot \vert \cdot)$ 
denotes the likelihood probability of symbol estimation for deep receivers. 
The deep receiver training objective can be described by

\begin{equation}
  \arg \underset{\theta}{\min }\left\{\mathcal{L}(\mathcal{T} ; \theta)=-\sum_{\left(\mathbf{y}_i, \mathbf{s}_i\right) \subset \mathcal{T}} \log \hat{P}_\theta\left(\mathbf{s}_i \mid \mathbf{y}_i\right)\right\} \text{.}
\end{equation}

The deep receiver, trained using $\mathcal{T}$, is used to decode the symbols in $\mathcal{V}$. 
For the $i$-th received symbol $\mathbf{y}_i$, the decoded result is expressed as $\hat{\mathbf{s}}_i\left(\mathbf{y}_i ;\theta\right)$. 
The performance metric of the deep receiver is the symbol error rate (SER), which is defined as 

\begin{equation}
  S E R(\theta)=\frac{1}{L_{\text{info}}} 
  \sum_{i=L_{\text{pilot}}+1}^L 
  \operatorname{Pr}\left(\hat{\mathbf{s}}_i\left(\mathbf{y}_i ; \theta\right) \neq \mathbf{s}_i\right) \text{.}
\end{equation}

\subsection{Modus Operandi of the Malicious User}

\begin{figure*}[!t] 
  \centering 
  \includegraphics[width=\textwidth]{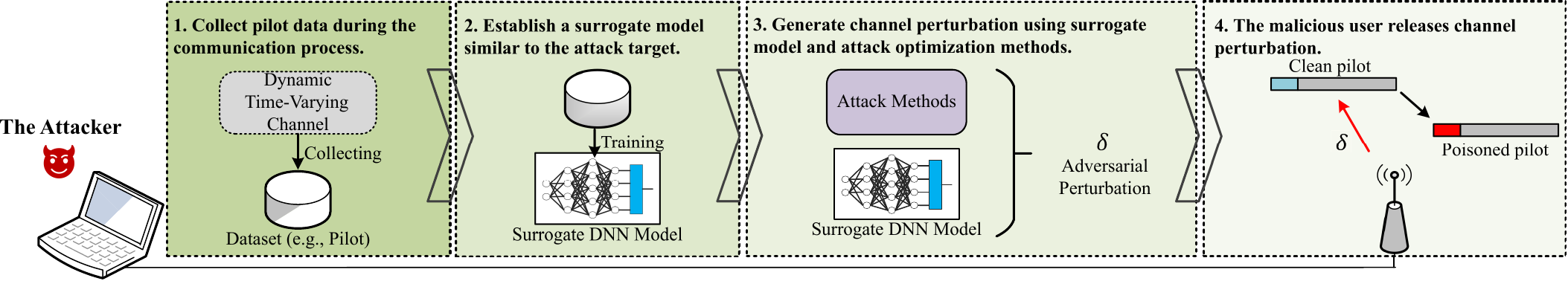} 
  \caption{Modus operandi of the malicious user.}
  \label{fig_4} 
\end{figure*}

For the considered system, as mentioned earlier,
there is a malicious user, which aims to corrupt the information
receiving and decoding of the legitimate receiver, 
by poisoning on the pilot data transmission from the transmitter to the receiver.
This is similar to \cite{sagduyu2019adversarial,pang2021accumulative}.
In particular, the attack process of the malicious user is shown in Fig. 4 and
summarised as below.

\begin{enumerate}[label=Step \arabic*:, leftmargin=*, widest=4]
  \item 
  Since the pilot pattern is fixed during transmission, the malicious user can collect pilot data during the communication process between the legitimate transceiver.
  \item The accumulated pilot data is employed to train a surrogate model 
  that is analogous to the attack target, specifically the authorised deep receiver.
  \item The malicious user generates the optimal perturbation based on the surrogate model and the transferability of the attack.
  \item The malicious user injects the channel perturbation.
  The deep receiver will gradually be poisoned until the model fails when it receives the perturbed pilot data used to train the model.
\end{enumerate}

In principle, a poisoning attack perpetrated 
by a malicious user can be conceptualized as a perturbation injection process.
In particular, the perturbation is defined as a vector in the complex space of receiver inputs, denoted as $\delta \in \mathbb{C}^{N_{rx}}$,
with a poisoning process represented by $\mathcal{P}(\cdot)$. 
For the $i$-th received symbol $\mathbf{y}_i$, the corresponding poisoning perturbation is given by $\delta_i$,
and the corresponding poisoned received symbol is given by $\mathcal{P}\left(\mathbf{y}_i\right)=\mathbf{y}_i+\delta_i $.
Therefore, within the context of poisoning attacks on deep receivers, 
the primary challenge for malicious users is to design an optimal perturbation signal structure that maximizes the deep receiver's loss 
on subsequent information symbols or validation sets, 
which will be addressed in the following sections.

\section{Adversarial Machine Learning Theory}

Before introducing the proposed poisoning attack method for online deep receivers, 
we briefly discuss the theory of adversarial machine learning.
Specifically, we provide a overview of the threat model presented in Section III-A, 
including the attacker's goal, knowledge, and capability. 
These concepts and definitions facilitate a more comprehensive understanding of the proposed attack method for deep receivers. 
Subsequently, in Section III-B and Section III-C, we briefly elucidate the fundamental optimization issues associated with two distinct attack paradigms, 
namely evasion attacks and data poisoning attacks. 
Furthermore, we explain the differences and connections between these two attack paradigms. 
Finally, in Section III-D, we analyze the transferability of attack samples and discuss the methods to enhance their transferability.

\subsection{Threat Model}

\subsubsection{Attacker's Goal}
As discussed in \cite{demontis2019adversarial},\cite{cina2023wild},\cite{lin2021ml}, the objective of the attacker in adversarial scenarios 
can be categorised according to the form of security threats, including integrity attacks and availability attacks.

\begin{itemize}
  \item[$\bullet$] \textbf{Integrity attacks}: The attacker's goal is to tamper the integrity of the target. 
  Specifically, this implies that the attack samples generated by the attacker are only effective in certain parts of the target system, 
  while the remainder of the target system retains its original functionality. 
  
  \item[$\bullet$] \textbf{Availability attacks}: In contrast, availability attacks aim to disrupt the entire system, 
  making it unavailable to legitimate users.
  
\end{itemize}

The main difference between integrity attacks and availability attacks lies in the focus of the optimization objectives of the attack models. 
In this paper, the proposed attack method is an availability attack, 
namely the destruction of the usability of all functions of a deep receiver.

\subsubsection{Attacker's Knowledge}
The attacker's knowledge indicates the extent to which they are aware of the attack target.
This knowledge encompasses several dimensions, as outlined in \cite{demontis2019adversarial},\cite{cina2023wild}. 
These dimensions include: 
(i) The data utilized for training purposes.
(ii) The architectural design of the target model, the learning algorithms employed during training, 
and their associated parameters and hyperparameters.
(iii) The data comprising the test set. 
Based on these dimensions, two main attack scenarios can be defined:

\begin{itemize}
  \item[$\bullet$] \textbf{White-box attacks}: 
  The attacker has complete knowledge about the attack target. 
  In this context, the attacker will adapt the nature of the attack  
  to align with the specific characteristics of the target to achieve the most effective and impactful outcome.
  
  \item[$\bullet$] \textbf{Black-box attacks}: 
  Black-box attacks can be further categorized into two main types: 
  Transfer-based attacks and query attacks. In transfer-based attacks, the attacker has limited or no knowledge of the target model.
  The knowledge about the target encompasses the aforementioned dimensions (i), (ii), and (iii).
  In this setting, the attacker is limited to relying on the data they have collected to construct a surrogate model that approximates the target model. 
  This attack is then transferred to the target model by launching a white-box attack on the surrogate model.
  In a black-box query attacks, the attacker can query the target's output or confidence level to optimize the attack. 
  Currently, the majority of black-box attacks exploit the transferability for attack purposes 
  \cite{demontis2019adversarial},\cite{xie2019improving,huang2023t,liu2023transferable,chen2023rethinking}.
  
\end{itemize}

The discussion regarding the attacker's knowledge aims to define the scenarios in which attacks are deployed, 
particularly in more practical black-box attacks, which are the focus of this paper. 
Moreover, within the framework of black-box attacks, the transferability of attack samples holds particular significance. 
This will be addressed in greater detail in Section III-D of this paper.

\subsubsection{Attacker's Capability}


The attacker's capabilities determine the methods used to influence the attack target and the constraints for data manipulation. 
To avoid potential defense filtering mechanisms, the attacker typically imposes an upper bound of $\epsilon$ to the perturbation $\delta$ under the $p$-norm space.
From the perspective of attack methods\cite{demontis2019adversarial}, 
if the attacker can manipulate data from both the training and testing phases, 
which considered as causal attacks and is called data poisoning attacks.
If the attacker can only manipulate the data during the testing phase, this attack is considered exploratory and is called evasion attacks. 
The difference lies in the optimization objectives and implementation methods.
This paper focuses on adversarial poisoning attacks in black-box scenarios, 
which can be seen as a synthesis of evasion attacks and data poisoning attacks. 
The specific optimization goals and implementation forms of evasion attacks and data poisoning attacks are described in the following 
Section III-B and Section III-C, respectively.

\begin{figure*}[!t]
  \centering
  \subfloat[Evasion attacks]{\includegraphics[width=3.05 in]{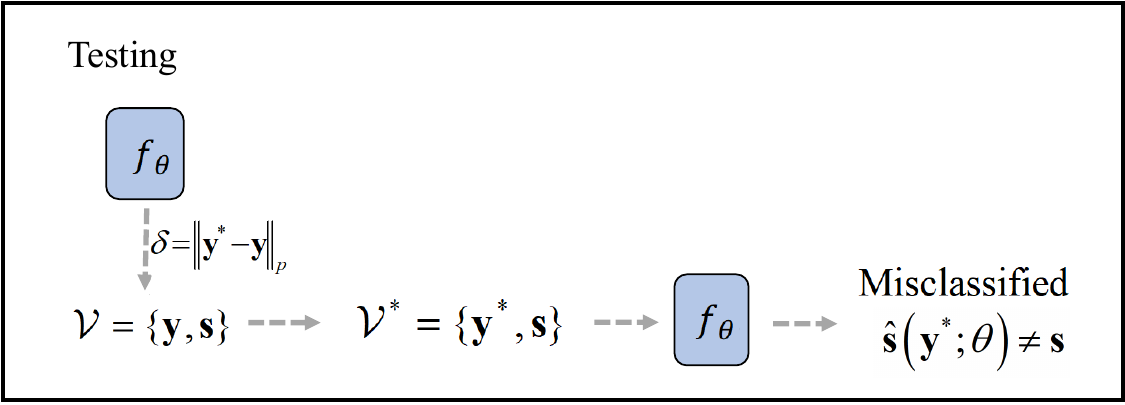}%
  \label{fig5_first_case}}
  \hfil
  \subfloat[Data poisoning attacks]{\includegraphics[width=3.86in]{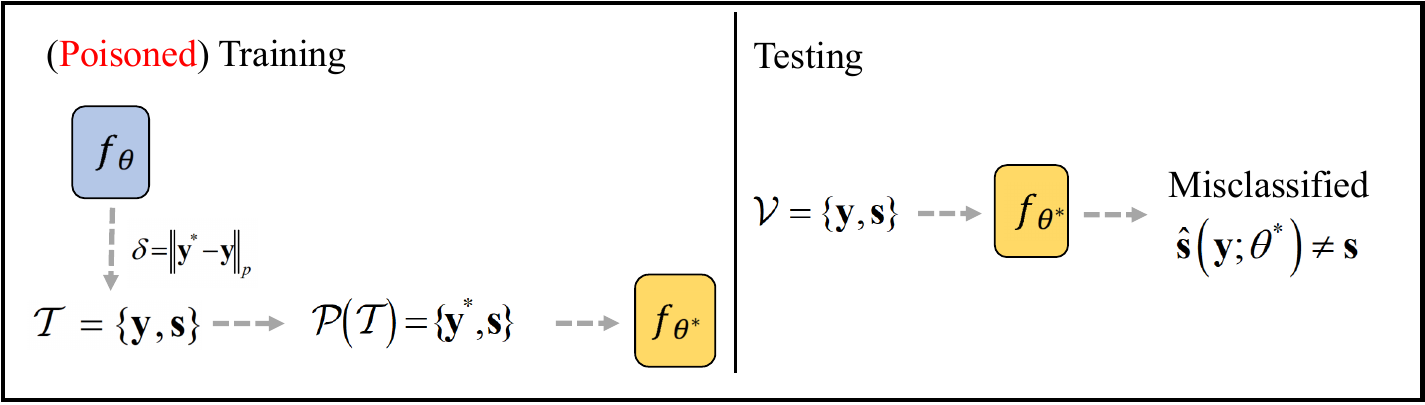}%
  \label{fig5_second_case}}
  \caption{The process of two attack paradigms.}
  \label{fig5}
\end{figure*}

\subsection{Evasion Attacks}

The evasion attacks executed by generating adversarial samples aimed at causing misclassification during the testing phase of the DNN.
This is achieved by identifying the network's vulnerabilities and applying small, strategically crafted perturbations to the input, as illustrated in Fig. 5(a). 
Gradient-based optimizers \cite{madry2017towards},\cite{dong2018boosting} are effective in determining these influential perturbations. 
Given a target model and input data, the gradient of the objective function is used to guide the application of minor perturbations, which maximizes the loss induced on the input. 
This can be formulated as a single-layer optimization problem. 
Specifically, let $\mathbf{y}$ represent the model input, $\mathbf{s}$ the corresponding labels, and $\mathbf{y}^{\prime}$ the adversarial samples generated after adding perturbation. 
The adversarial perturbation, denoted as $\delta=\left\|\mathbf{y}^{\prime }-\mathbf{y}\right\|_p$, is constrained by an upper bound $\epsilon>0$, with the optimal adversarial sample $\mathbf{y}^*$. 
The classifier is parameterized by $\theta$, and for classification tasks, the cross-entropy loss function $\mathcal{L}\left(\mathbf{y}^{\prime}, \mathbf{s}; f\theta\right)$ is employed. 
Thus, the generation of optimal adversarial samples can be framed as the following optimization problem

\begin{equation}
  \mathbf{y}^* \in \arg\underset{\mathbf{y}^{\prime}}{ \max } \mathcal{L}\left(\mathbf{y}^{\prime}, \mathbf{x} ; \theta\right) \text {, such that } \delta=\left\|\mathbf{y}^{\prime}-\mathbf{y}\right\|_p \leq \epsilon \text {. }
  \label{eq:4}
\end{equation}

\subsection{Data Poisoning Attacks}

The optimization goal of the data poisoning attacks is to poison the target with poisoned training data to degrade its test performance, 
which is shown in Fig. 5(b). 
Similar to the generation of adversarial samples, in data poisoning attacks, 
a perturbation $\delta$ is applied to each sample in the training set under the $p$-norm constraint, 
ensuring that the perturbation magnitude does not exceed an upper bound $\epsilon$.
Specifically, the test set $\mathcal{V}$ and the training set $\mathcal{T}$ are defined, 
as well as the poisoned training set $\mathcal{P}(\mathcal{T})$ after applying the perturbation to the data set. 
Furthermore, $\theta^*$ denotes the optimal poisoning parameter of the target classifier with respect to the parameter $\theta$. 
Thus, the poisoning attack can be modelled as a dual optimization problem as follows:

\begin{equation}
  \max _{\mathcal{P}} \mathcal{L}\left(\mathcal{V} ; \theta^*\right) \text {, and } \theta^* \in \arg \min \mathcal{L}(\mathcal{P}(\mathcal{T}) ; \theta){.}
  \label{eq:5}
\end{equation}

Herein, firstly, the inner layer optimization involves the standard model training process. 
In this process, the attacker uses the poisoned data to train the target model by minimizing the empirical loss $\mathcal{L}(\mathcal{P}(\mathcal{T}); \theta)$ 
to obtain the optimal poisoned parameter $\theta^*$. Secondly, based on the obtained $\theta^*$, 
the attacker maximizes the loss $\mathcal{L}(\mathcal{V}; \theta^*)$ on the test set. 
Note that solving this optimization problem directly is often very difficult, 
and it is more common practice to approximate this maximization process through gradient optimization of $\delta$.


\subsubsection{Adversarial Samples as Poisoning Attacks}
As previously stated in Section III-B and Section III-C, 
although both construction of evasion attacks and poisoning attacks 
can be framed as gradient-based optimization problems, 
the goals achieved by constructing perturbations for these two types of attacks are different. 
Recently, however, researchers have discovered that adversarial samples are also highly effective for poisoning DNNs, 
a phenomenon known as \textit{Adversarial Poisoning} \cite{fowl2021adversarial},\cite{sandoval2024can}. 
In this case, the poisoning attack optimization problem (\ref*{eq:5}) can also be uniformly expressed 
in the form of the adversarial sample attack optimization problem (\ref*{eq:4}) in Section III-B. 
\cite{fowl2021adversarial} provides a method for creating adversarial poisoning samples to obtain optimal poisoning results. 
Additionally, \cite{oldewage2023adversarial} shows that adversarial poisoning attacks can also cause serious harm to the meta-learner in a white-box attack setting. 
Compared with the dual optimization process of the poisoning attack method in (\ref*{eq:5}), 
using adversarial samples as poisoning attack samples is more convenient and practically feasible. 
This is also the basis for the poisoning attack method proposed in this paper.

\subsection{Transferability of Attacks}
\subsubsection{Why Can Attacks Be Transferred?}
Attack transferability means that the attack generated for a model may be equally effective for other models that have not seen that attack before. 
This phenomenon has been observed and demonstrated in \cite{liu2016delving}. 
The proposed attack method is a black-box attack. 
Therefore, it is crucial to analyze the source of the attack portability. 

\cite{demontis2019adversarial} presented an upper bound on the loss when black-box transfer occurs. 
Define $f_{\varphi}$ as the surrogate model with parameter $\varphi$, 
$f_\theta$ as the target model with parameter $\theta$, and $\mathcal{L}(\mathbf{y}, \mathbf{s}, \varphi)$ as the loss of the input $\mathbf{y}$ 
in $f_{\varphi}$ against the label $\mathbf{s}$. 
In consideration of the transferability of evasion attacks (poisoning attacks also take the same form), 
the optimal adversarial sample, denoted as $\mathbf{y}^*$ is obtained by solving (\ref*{eq:4}) on $f_{\varphi}$, 
with the corresponding optimal perturbation, denoted as $\delta^*$.
To illustrate, consider the sphere space with $p=2$ and radius denoted as $\epsilon$. 
The optimal adversarial perturbation obtained on $f_{\varphi}$ can be expressed in (\ref{eq:6}) as follows:

\begin{equation}
  \label{eq:6}
  \delta^*=\epsilon \frac{\nabla_{\mathbf{y}} \mathcal{L}(\mathbf{y}, \mathbf{s}, \varphi)}{\left\|\nabla_{\mathbf{y}} \mathcal{L}(\mathbf{y}, \mathbf{s}, \varphi)\right\|_2}.
\end{equation}

As a result, the loss of $\mathbf{y}$ on the target model, denoted as $\mathcal{L}(\mathbf{y}, \mathbf{s}, \theta)$. 
Define $\Delta \mathcal{L}$ as the increase in loss of the input $\mathbf{y}^*$ compared to the input $\mathbf{y}$ on the target model. 
The upper bound of  $\Delta \mathcal{L}$ on the target model can be described by 

\begin{equation}
  \label{eq:7}
  \Delta \mathcal{L}=\epsilon \frac{\nabla_{\mathbf{y}} \mathcal{L}(\mathbf{y}, \mathbf{s}, \varphi)^{\top}}{\left\|\nabla_{\mathbf{y}} \mathcal{L}(\mathbf{y}, \mathbf{s}, \varphi)\right\|_2} \nabla_{\mathbf{y}} \mathcal{L}(\mathbf{y}, \mathbf{s}, \theta) \leq \epsilon\left\|\nabla_{\mathbf{y}} \mathcal{L}(\mathbf{y}, \mathbf{s}, \theta)\right\|_2.
\end{equation}

The left-hand side of the inequality in (\ref{eq:7}) represents the loss in the black-box attack scenario, 
while the right-hand side represents the loss in the white-box attack scenario. 
In the white-box attack scenario, i.e., when $f_{\varphi} = f_\theta$, the inequalities in (\ref{eq:7}) becomes an equality.
At this point, the attack achieves its upper bound and has optimal effectiveness.

Thus, the effectiveness of transferring an attack sample from the surrogate model to the target model 
is influenced by two factors: \textit{The intrinsic adversarial vulnerability of the target model} (right-hand side of the inequality in (\ref{eq:7})) 
and \textit{the complexity of the surrogate model used to optimize the attack} (left-hand side of the inequality in (\ref{eq:7})).
The right-hand side of the inequality in (\ref{eq:7}) shows that a more vulnerable target model has a larger upper bound on the loss, 
represented by $\epsilon\left|\left|\nabla_{\mathbf{y}} \mathcal{L}(\mathbf{y}, \mathbf{s}, \theta)\right|\right|_2$. 
Here, the intrinsic complexity of the model measures the learning algorithm's ability to fit the training data. 
More complex models, like those without regularization or those prone to overfitting,
have more intricate parameter spaces and rugged loss landscapes, making them sensitive to input perturbations.
For robust models with smaller upper loss bounds, a successful attack requires a higher perturbation limit, 
reducing the likelihood of bypassing the system's monitoring.
This demonstrates the impact of the intrinsic adversarial vulnerability of the target model on transferability of attacks.

The complexity of the surrogate model used to optimize the attack depends on two main factors: 
The gradient alignment between the surrogate and the target, 
and the variance magnitude of the surrogate model's loss function.
These factors are particularly relevant for the left-hand side of the inequality in (\ref{eq:7}).
When the surrogate has better gradient alignment with the target, 
such as higher gradient cosine similarity \cite{demontis2019adversarial},
the attack samples from the surrogate model exhibit better transferability. 
This is reflected in $\nabla_{\mathbf{y}} \mathcal{L}(\mathbf{y}, \mathbf{s}, \varphi)^{\top} \nabla_{\mathbf{y}} \mathcal{L}(\mathbf{y}, \mathbf{s}, \theta)$, 
as shown on the left-hand side of the inequality in (\ref{eq:7}). 
Additionally, a surrogate model with low variance leads to a more stable optimization process, 
producing attack samples effective across different target models. 
In contrast, a large variance leads to an unstable optimization process, leading to attack samples may not match the target model, resulting in failure. 
Intuitively, on the left-hand side of the inequality in (\ref{eq:7}), high variance of the loss function increases the corresponding denominator term 
$\left|\left|\nabla_{\mathbf{y}} \mathcal{L}(\mathbf{y}, \mathbf{s}, \varphi)\right|\right|_2$, 
which results in smaller upper bounds on the achievability of the transfer.

\subsubsection{Related Work of Transfer-based Attacks}
In summary, transfer-based attacks can be approached from two perspectives: 
The inherent adversarial vulnerability of the target model and the complexity of optimizing the surrogate model.

\begin{itemize}
  \item[$\bullet$] For the former, certain assumptions about the target model are typically necessary to model the attack objective. 
  For example, \cite{chen2023rethinking} assumed unstable common weakness in the model ensemble.
  In this scenario, stronger constraints can be applied to the optimization target, such as sharpness-aware minimization\cite{chen2023rethinking} 
  or momentum methods \cite{dong2018boosting}, to make the transfer attack more effective.
  However, the intrinsic adversarial vulnerability of unknown target is often difficult to identified directly.
  
  \item[$\bullet$] 
  For the latter, approaches include using ensembled surrogate model \cite{liu2016delving}, 
  self-ensembled method \cite{huang2023t}, and data augmentation \cite{huang2023t,xie2019improving}. 
  \cite{liu2023transferable} suggested alternating different training paradigms 
  (e.g., unsupervised and self-supervised) to enhance transferability in poisoning attacks.
  These methods aim to build more generalized and robust surrogate models, 
  which are then optimized to obtain more effective and transferable attack samples. 
  
\end{itemize}

\section{A poisoning attack framework \\ for online deep receivers}

\begin{figure*}[!t] 
  \centering 
  \includegraphics[width=\textwidth]{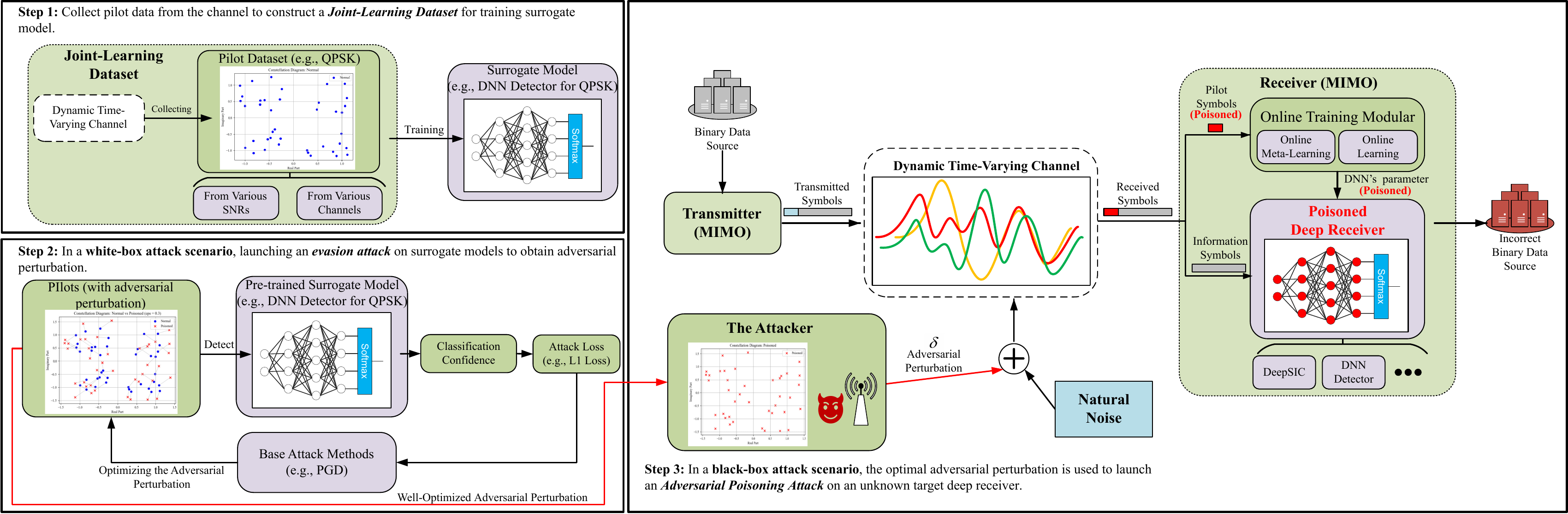} 
  \caption{A poisoning attack framework for online deep receivers.}
  \label{fig_6} 
\end{figure*}

As mentioned earlier, in the scenarios discussed in this paper, 
legitimate deep receivers adapt to fast-varying wireless channels by utilizing pilot and updating model parameters through online learning.
However, this online update mechanism, designed to adapt to local channel variations, is vulnerable to sample input poisoning.

In the attack framework of the malicious user, the legitimate online deep receiver becomes the target model,
which is updated online to adapt to local channel variations for better performance.
However, this process is inherently not robust and can result in overfitting \cite{shi2021overcoming}.
From this point onwards, the attack target of this paper is similar to \cite{perry2020lethean}, 
which causes catastrophic forgetting of the target by constructing poisoned samples.
According to the analysis in Section III-D, 
this overfitting is the source of the target model's intrinsic adversarial vulnerability.
This implies that a malicious user can effectively attack the target model by optimizing a surrogate model 
and generating corresponding adversarial perturbations.

Therefore, this section proposes a poisoning attack framework targeting online deep receivers. 
The core idea is to poison the model training and updating phases, resulting in a degraded model after a certain period.
Specifically, based on the behavioral patterns of malicious user described in Section II-C, 
the poisoning attack framework and the method for generating the poisoning attack samples can be summarized in below 3 steps, as illustrated in Fig. 7.

\begin{enumerate}[label=Step \arabic*:, leftmargin=*, widest=4]
  \item The malicious user collects communication data (e.g., pilot) from the wireless channel and produces a joint learning dataset. 
  This dataset is used to train a surrogate model for attacking the target (i.e., the legitimate deep receiver).
  \item 
  The malicious user optimization solution (\ref*{eq:4}) generates an adversarial perturbation based on the surrogate model.
  \item The malicious user injects the adversarial perturbation onto the channel, causing the deep receiver to receive the poisoned pilot. 
  Consequently, the deep receiver undergoes online learning with the perturbed pilot, resulting in the model being poisoned and deactivated.
\end{enumerate}

The specific details of Steps 1 and 2 are detailed in the following Sections IV-A to IV-C.

\subsection{Surrogate Model Selection}
As discussed in Section III-D, black-box attacks rely on the gradient alignment between the surrogate and target models to ensure generalization. 
To achieve this, it is crucial to avoid using surrogate models that are too specialized for specific tasks. 
This strategy improves compatibility with different deep receivers, enhancing the attack's effectiveness. 
Therefore, this paper employs a generic DNN architecture, such as feedforward neural networks, as the surrogate model, as shown in Step 1 of Fig. 6.


\subsection{Joint Learning for Training of Surrogate Models}
From the perspective of a malicious user, 
it is essential to select a suitable surrogate model architecture similar to the target model 
and to effectively train and optimize the surrogate model.
In the proposed attack framework, joint learning is used to train the surrogate model. 
This approach utilizes data collected under various channel conditions to train the DNN, enabling it to adapt to dynamic channels
\cite{o2017introduction,raviv2024adaptive}.
Unlike legitimate communicating parties that have to use online learning methods to adapt to time-varying wireless channels, 
the malicious user can pre-collect large datasets to train a robust model. 
Furthermore, the use of joint learning, instead of using channel state information or online learning \cite{raviv2024adaptive}, 
addresses the issue of attack generality at the data level, avoiding over-specialized surrogate models in black-box attack transfers, 
as discussed in Section IV-A.
Joint learning also satisfies the data volume requirements of deep learning, 
making the surrogate model more robust than the target model, which uses online learning.
This results in attack patterns with more stable and effective attack results.
Conversely, the robust learning process of the target model mitigates the impact of poisoning,
as demonstrated by pilot size adjustments in Section V-E. 

Specifically, as shown in Step 2 of Fig. 6, the malicious user employs joint learning to train a DNN (i.e., the surrogate model) 
utilizing a large amount of communication data (e.g., pilot) amassed from legitimate transceivers.
The DNN learns a mapping applicable to most channel states, reflecting the input-output relationship of the target deep receivers. 
The training data includes two types: 
communication data from different channel distributions and data from varying signal-to-noise ratio (SNR) conditions within the same channel.
Finally, to generate effective adversarial poisoning samples, suitable attack generation methods must be considered, as detailed in Section IV-C.

\subsection{Adversarial Poisoning Attack Samples Generation}

\begin{algorithm}
  \caption{Sample Generation Method for Poisoning Attack Based on PGD Algorithm.}\label{alg:alg1}
  \begin{algorithmic}
    \STATE
    \STATE \textbf{Input:} Pilot dataset $\mathcal{T}_\text{pilot} = \{\mathbf{y}, \mathbf{s}\}$, 
    model parameter $\varphi$, PGD iteration count $Q$, 
    perturbation step size $\gamma$, loss function $\mathcal{L}(\mathcal{T}; \varphi)$, 
    input data perturbation bounds $I_{max}/I_{min}$, perturbation limit $\epsilon$
    \STATE \textbf{Output:} Poisoned dataset $\mathcal{P}(\mathcal{T}) \equiv\left\{\mathbf{y}^*, \mathbf{s}\right\}$

    \STATE Initialize poisoned dataset: $\mathcal{P}(\mathcal{T}) = \emptyset$
    
    \FOR {each symbol $\mathbf{y}$ in the pilot dataset $T_\text{pilot} = \{\mathbf{y}, \mathbf{s}\}$}
      \STATE $\delta \sim U(-\epsilon, \epsilon)$
      \STATE $\mathbf{y}^* = \text{clip}(\mathbf{y} + \delta, I_{min}, I_{max})$
      
      \FOR {$i \in \{1, \ldots, Q\}$}
        \STATE $\delta \leftarrow \text{sgn}\left(\nabla_{\mathbf{y}} \mathcal{L}\left(\mathbf{y^*}; s; f_{\varphi}\right)\right)$
        \STATE $\mathbf{y}^* \leftarrow \text{clip}(\mathbf{y} + \gamma \delta, I_{min}, I_{max})$
        \STATE $\mathbf{y}^* \leftarrow \mathbf{y} + \text{clip}(\mathbf{y}^* - \mathbf{y}, -\epsilon, \epsilon)$
      \ENDFOR
      
      \STATE $\mathcal{P}(\mathcal{T}) = \mathcal{P}(\mathcal{T}) \cup \{\mathbf{y}^*, \mathbf{s}\}$
    \ENDFOR
  \end{algorithmic}
  \label{alg1}
\end{algorithm}

Once the optimized surrogate model has been obtained, the malicious user generates the necessary poisoning attack samples 
in order to execute the attack. 
As illustrated in Step 2 of Fig. 6, we employ the projected gradient descent (PGD) algorithm \cite{madry2017towards}, 
to generate the adversarial poisoning attack perturbations discussed in Section III-C. 
The whole algorithmic flow of generating perturbations is shown in Algorithm 1,
and summarised as below.

\begin{enumerate}[label=Step \arabic*:, leftmargin=*, widest=4]
  \item Obtain the pilot dataset $\mathcal{T}_{\text{pilot}} \equiv \{\mathbf{y}, \mathbf{s}\}$. 
  Then, within an interval $[-\epsilon, +\epsilon]$ defined by the upper bound of the $p$-norm constraints, 
  use uniformly distributed sampling to generate a randomly initialized perturbation vector $\delta$.

  \item Superimpose the $\delta$  on the pilot data. 
  The perturbed data should then be fed into the surrogate model $f_\varphi$ to calculate the attack loss.
  
  \item 
  $\delta$ updates in the gradient direction of loss with step size $\gamma$,
  and $\mathbf{y} + \gamma \delta$ remains within the specified upper bound $I_{max}$, and lower bound $I_{min}$.
  
  \item 
  The Step 2 and Step 3 iteratively repeat the $Q$ rounds to obtain poisoned pilot data with the optimal attack perturbation.

\end{enumerate}

Once the iteration is complete, the optimal perturbation will be applied to the current block's pilot to generate the poisoned pilot dataset, denoted as $\mathcal{P}(\mathcal{T})$. 
This dataset will then be received by the deep receiver and poisoned during the training and updating of the model, as illustrated in Step 3 of Fig. 6.

\section{NUMERICAL EVALUATIONS}
In this section, we numerically evaluate the proposed poisoning attack aimed at disrupting the deep receiver's online adaptation process. 
First, we outline the parameter settings used in the simulations, including channel models, deep receivers, online training methods, 
and the poisoning attack method (Sections V-A to V-D). 
Next, simulation results are presented for deep receivers under the following conditions: linear and nonlinear time-varying synthetic channels, 
a linear static synthetic channel, and the time-varying COST 2100 channel (Section V-E). 
Lastly, the experimental results across all four channel settings are summarized and discussed in Section V-F.


\subsection{Evaluated Channel Models}

The deep receiver operates on a discrete memoryless MIMO channel.
The number of transmitting and receiving antennas is set to $N_{t x}$$=$$N_{r x}$$=$4. 
The experimental evaluation channel model comprises synthetic channels \cite{raviv2023online} 
and COST 2100 channels \cite{liu2012cost}. 
Fig. 7 illustrates the four channel tapping coefficients for a randomly selected user in a multiuser system over 100 blocks of data transmission.
In the context of linear channels, the input-output relationship is expressed by the (\ref{eq:1}) given in Section II-A. 
In the context of a nonlinear channel model \cite{raviv2023online}, the input-output relationship is represented by 

\begin{figure*}[!t]
  \centering
  \subfloat[Synthetic gaussian channel]{\includegraphics[width=3.in]{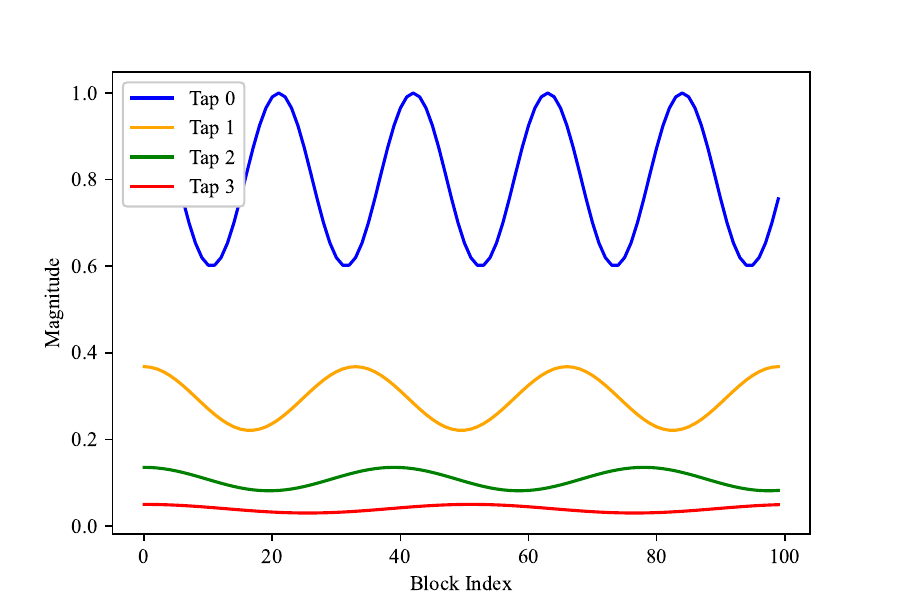}%
  \label{fig_first_case}}
  \hfil
  \subfloat[COST 2100 channel]{\includegraphics[width=3.in]{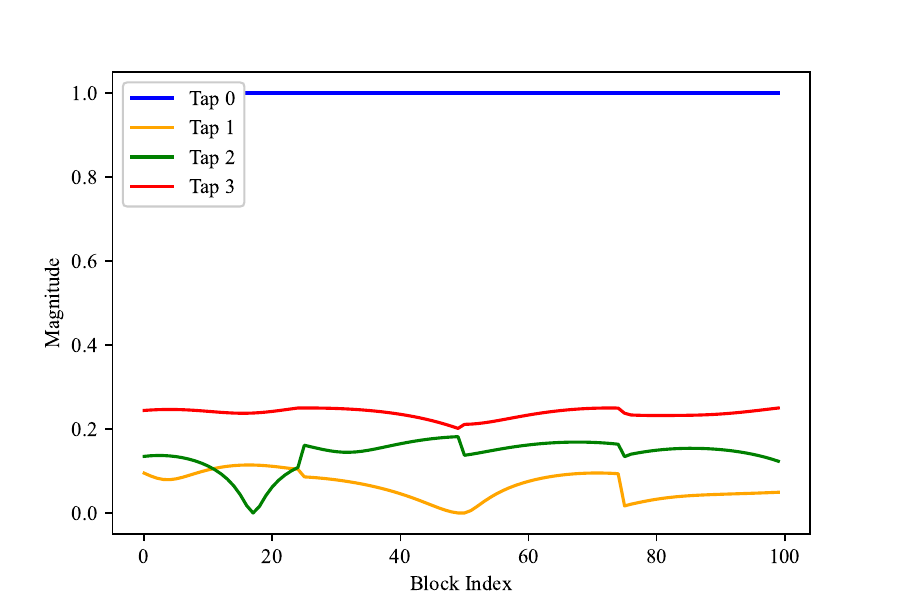}%
  \label{fig_second_case}}
  \caption{Time-varying channel tap coefficient variation.}
  \label{fig7}
\end{figure*}

\begin{equation}
  \mathbf{y}=\tanh (k(\mathbf{H x}+\mathbf{w})),
\end{equation} 
where the $\tanh (\boldsymbol{\cdot})$ function is used to simulate the non-linear variations in the transceiver process 
due to non-ideal hardware, with the parameter $k=0.5$.

\subsection{Evaluated Deep Receivers}


We consider three deep receiver architectures in the evaluated network architecture: Namely, 
the model-based deep receiver DeepSIC \cite{shlezinger2020deepsic}, the black-box DNN detector \cite{raviv2024adaptive},\cite{raviv2023data}, 
and the ResNet detector designed based on DeepRX \cite{honkala2021deeprx}.The relevant details are as follows:

\begin{itemize}
  \item[$\bullet$] \textbf{DeepSIC}: 
  DeepSIC unfolds the traditional SIC iterative process and replaces each iteration with a sub-neural network to improve performance. 
  Each sub-neural network enhances the reliability of the current estimation using received symbols and the output confidence from the previous iteration.
  This design enables DeepSIC to achieve high reliability even with limited training data\cite{shlezinger2020deepsic,raviv2023online}. 
  In this paper, DeepSIC includes 3 iterations, resulting in a total of $3 \times N_{t x}$ sub-networks.
  Each sub-network is a two-layer fully connected layer network. 
  The first layer has a dimension of $(N_{r x} + N_{t x} - 1) \times 64$, 
  and the second layer has a dimension of $64 \times |S|$, where $|S|$ is the size of the set of symbols to be transmitted. 
  For example, $|S| = 4$ when using QPSK transmission. 
  The activation function employed in the initial layer of each subnetwork is ReLU, 
  while the second layer utilizes softmax classification.
  \item[$\bullet$] \textbf{Black-box DNN Detector}: 
  The black-box DNN architecture consists of 4 fully-connected layers and a softmax classification header. 
  The dimensions of the layers are $N_{r x} \times 60,60 \times 60,60 \times 60$ and $60 \times|S|^{N_{tx}}$, respectively.
  \item[$\bullet$] \textbf{ResNet Detector}: 
  The ResNet detector employed in this paper comprises 10 layers of residual blocks, 
  each containing 2 convolutional layers with 3$\times$3 kernels, 
  one-pixel padding on both sides, and no bias terms. 
  A ReLU activation function is used between the layers, and each convolutional layer is followed by 2D batch normalization.
\end{itemize}

\subsection{Online Training Methods}

\begin{table}[ht]
  \caption{The parameter configuration of the Online Training Methods}
  \centering
  \small 
  \begin{tabular}{@{}>{\centering\arraybackslash}p{4cm} >{\centering\arraybackslash}p{3cm}@{}} 
    \toprule 
    \textbf{Parameters} & \textbf{Values} \\
    \midrule 
    $L_\text{pilot}$ & 200 \\
    $L_\text{info}$ & 50000 \\
    $\eta$ & $5 \times 10^{-3}$ \\
    $\eta_{\text{meta}}$ & $0.01 $ \\
    Epochs & 300 \\
    Optimizer & adam \\
    \bottomrule 
  \end{tabular}
  \label{tb2} 
\end{table}

The objective of this paper is to present an attack strategy, which attacks online adaptation of the deep receivers. 
Consequently, the focus is on the deep receiver's online training, with the parameter configurations illustrated in Table I.
The parameter settings for online training come from \cite{raviv2024adaptive,raviv2023data}.
Upon receipt of a data block, the deep receiver is only able to utilize a subset of the data block for training, 
like pilot symbols, specifically $L_{\text {pilot}} = 200$.
Then the deep reciver predicts the subsequent $L_{\text {info}} = 50000$ symbols. 
In the experiment, a total of 100 data blocks were transmitted, 
with the transmitted data being QPSK modulated, i.e., the user-transmitted symbols $\mathbf{s}$ were mapped to the set 
$C=\left\{\left( \pm \frac{1}{\sqrt{2}}, \pm \frac{1}{\sqrt{2}}\right)\right\}^4$. 
Furthermore, the deep receivers are trained using the adam optimizer.
The training epochs are 300. 
The initial learning rate $\eta$ is set to 5 $\times 10^{-3}$  for ResNet detector, black-box DNN detector and DeepSIC. 
The meta learning rate $\eta_{\text{meta}}$ is set to 0.01 for Meta-DeepSIC.
Additionally, online training is performed for different architectural receivers, 
which were implemented in the following two cases:
\begin{itemize}
  \item[$\bullet$] \textbf{Online learning}: 
  Based on the adaptation from the previous data block, the deep receiver trains 
  and updates the current model using the limited pilot symbols received in the current data block.
  \item[$\bullet$] \textbf{Online meta-learning}: 
  According to \cite{raviv2023online}, \cite{finn2017model}, the meta-learning is employed to facilitate adaptation to the channel.
  Specifically, the pilot data from 5 data blocks is accumulated, and then meta-learning is 
  performed on the this data to obtain the meta-learning weights for the deep receiver. 
  Subsequently, the weights are then employed in an online learning process involving the pilot data of the current block.
\end{itemize}

Unless otherwise stated,
the deep receivers are trained using online learning, 
including black-box DNN detector, ResNet detector, and DeepSIC. 
Only Meta-DeepSIC is trained using online meta-learning methods.

\subsection{Attacker's Configuration}

\begin{table}[ht]
  \caption{The parameter configuration of the Attack Methods}
  \centering
  \small 
  \begin{tabular}{@{}>{\centering\arraybackslash}p{4cm} >{\centering\arraybackslash}p{3cm}@{}} 
    \toprule 
    \textbf{Parameters} & \textbf{Values} \\
    \midrule 

      $ p $ &	2 \\
      $ \epsilon $  &	0.3 \\
      $ \gamma $ &	0.01 \\
      $ Q $ &	250 \\
      $ L_\text{sur} $	& 5000\\
      $ SNR_\text{sur} $ &	\{2,4,6,8,10,12,14,16\}   \\
      
    \bottomrule 
  \end{tabular}
  \label{tb3} 
\end{table}

The attack samples are designed based on the gradient of the surrogate model, 
with joint learning performed using the collected pilot dataset. 
The black-box DNN detector, presented in Section V-B, is employed as the surrogate model 
and contains three times the number of parameters compared to a single sub-network of DeepSIC. 
The joint learning parameters are based on \cite{raviv2024adaptive}. 
In particular, the linear time-varying synthetic channel model is employed to generate the channel data for joint learning.
Training data is generated under the condition of an SNR of 2 dB to 16 dB with an interval of 2 dB, denoted as $SNR_\text{sur} $.
In this manner, $L_\text{sur}=5000$ training pilot symbols are generated at each SNR value, 
and the surrogate model is trained in accordance with this procedure. 

Moreover, the poisoning samples are iteratively optimized using the PGD algorithm, following the settings in \cite{fowl2021adversarial}, 
where the adversarial poisoning samples are most effective under this specific parameter configuration.
The iteration step size is set to $\gamma = 0.01$, 
the iterations of PGD is $Q = 250$, and the perturbation upper bound under the $p = 2$ norm is $\epsilon = 0.3$. 
The maximum and minimum values of the received symbol magnitude are denoted as $I_{max}$ and $I_{min}$, respectively, 
and $I_{max} =  max \{ \|\mathbf{y}\| \} = -I_{min} $.
Fig. 8 shows the real and imaginary parts of the original and poisoned received symbols, while Fig. 9 presents these symbols in a constellation diagram.

\begin{figure}[!h] 
  \centering 
  \includegraphics[width=0.45\textwidth]{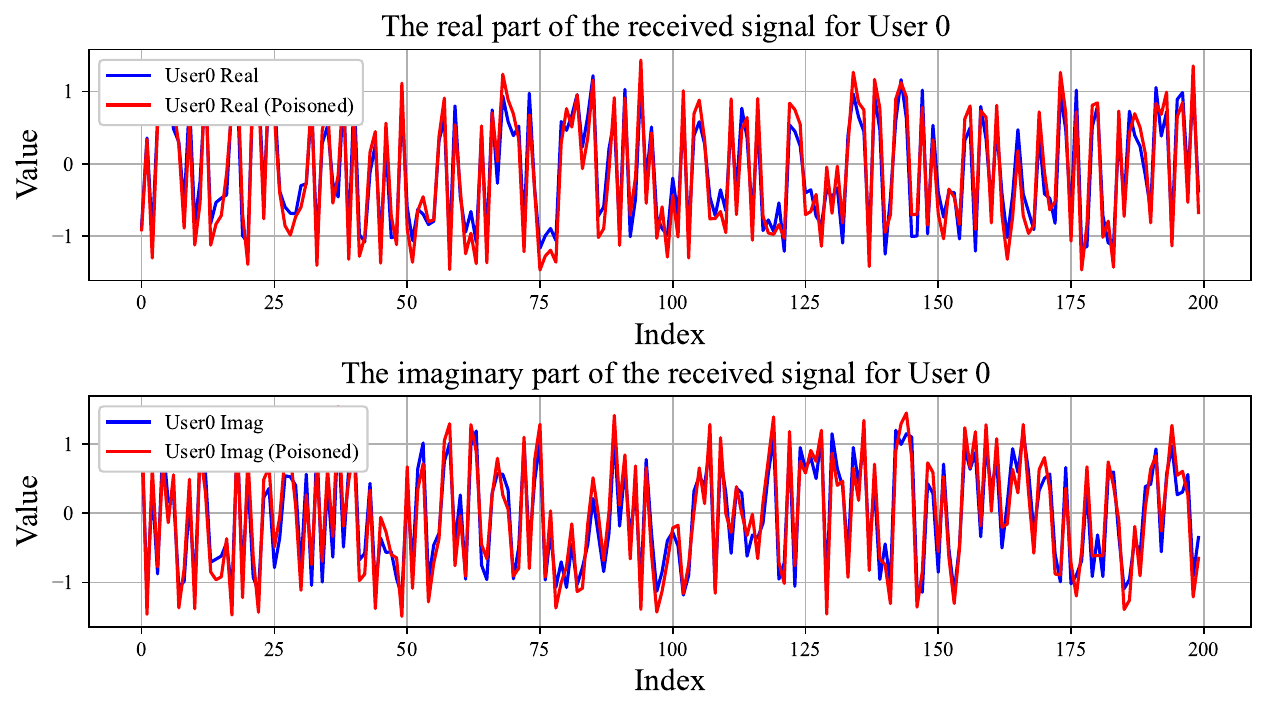} 
  \caption{Comparison of time domain for poisoned received symbols and normal received symbols.}
  \label{fig_8} 
\end{figure}

\begin{figure}[!h] 
  \centering 
  \includegraphics[width=0.4\textwidth]{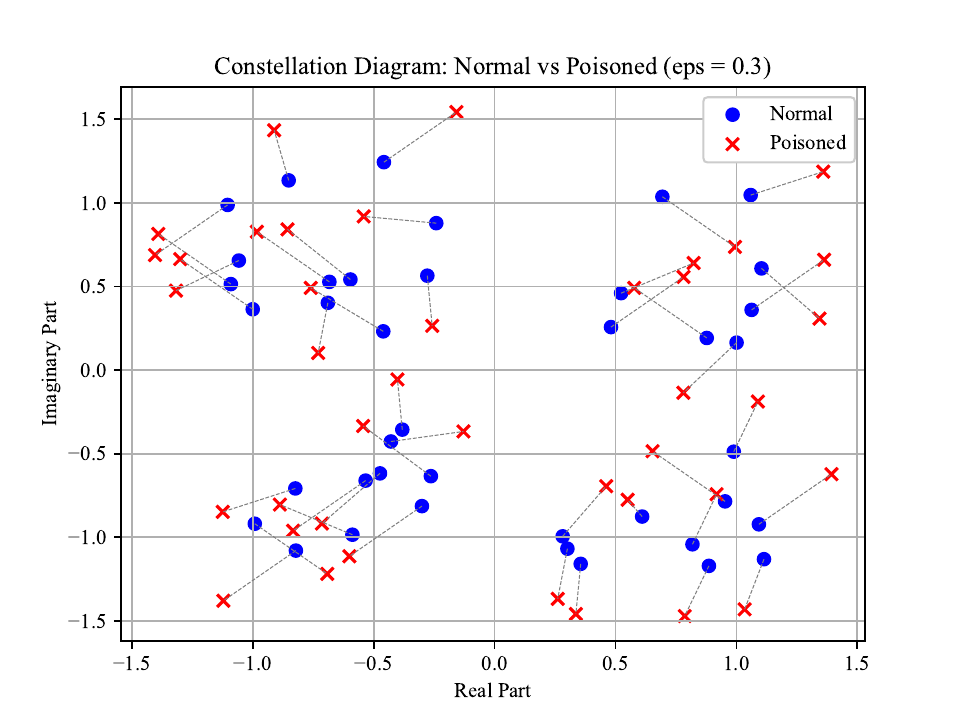} 
  \caption{Comparison of constellation diagrams for poisoned received symbols and normal received symbols.}
  \label{fig_9} 
\end{figure}

\subsection{Numerical Results under Four Channel Models}
This section presents the experimental results evaluating the proposed poisoning attack across various channel models.
The method's effectiveness is tested on linear and nonlinear time-varying synthetic channels, linear static synthetic channels, and time-varying COST 2100 channels.
For the black-box DNN detector, a white-box poisoning attack is launched against this architecture receiver since its architecture is the same as the surrogate model. 
In the case of the ResNet detector, transfer poisoning attacks are implemented in black-box scenarios. 
Since DeepSIC contains multiple sub-networks, direct gradient information cannot be used for designing poisoning samples. 
Therefore, a transfer-based poisoning attack is employed, anticipating that perturbations designed on the surrogate model will transfer to DeepSIC.


\subsubsection{Linear Time-varying Synthesis Channel Results}

\begin{figure*}[!t]
  \centering
  \subfloat[\textrm{SER vs. block index, SNR = 14 dB}]{\includegraphics[width=2.95in]{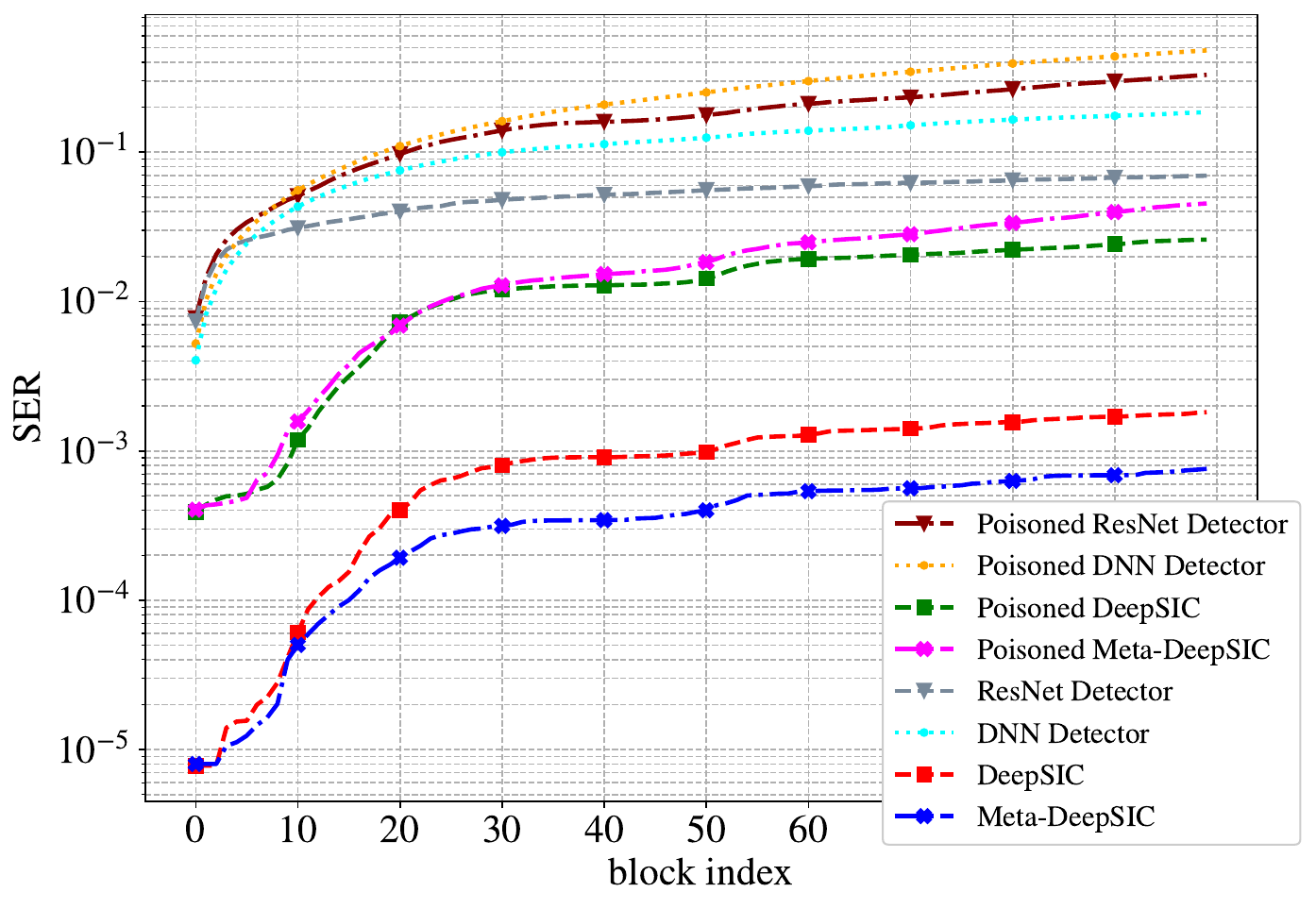}%
  \label{fig10_first_case}}
  \hfil
  \subfloat[\textrm{SER vs. SNR}]{\includegraphics[width=3.in]{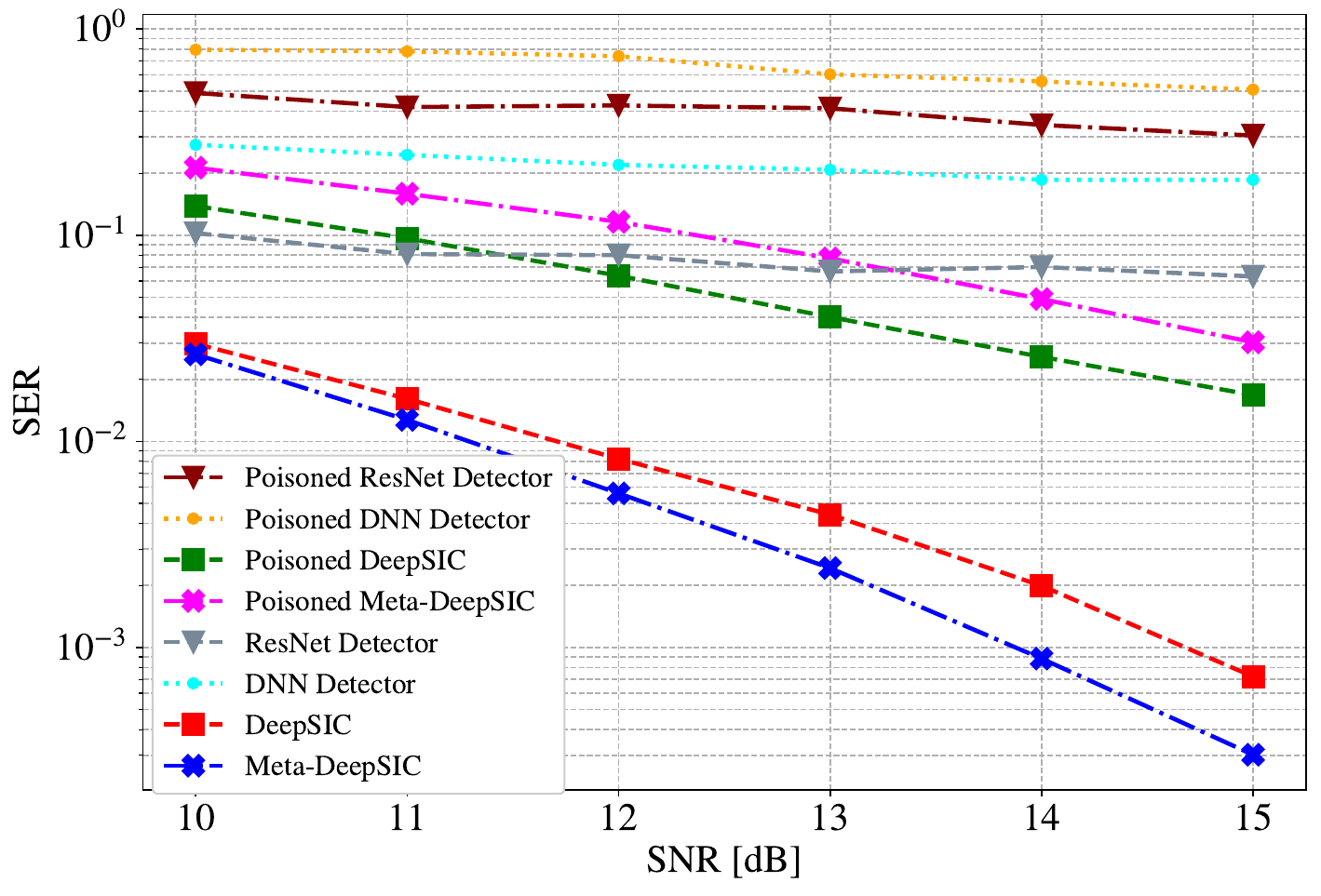}%
  \label{fig10_second_case}}
  \caption{Experimental results under Linear Time-varying Synthesis Channel.}
  \label{fig10}
\end{figure*}

Fig. 10(a) shows the results of 100 data transmission blocks, 
with the average cumulative SER calculated block-wise over five repetitions at an SNR of 14 dB. 
Meta-DeepSIC, combining model-based method with meta-learning, 
better captures the time-varying wireless channel characteristics in the absence of a poisoning attack, 
achieveing superior SER performance compared to the black-box DNN detector, ResNet detector and DeepSIC. 
However, Meta-DeepSIC is more susceptible to poisoning attacks than the DeepSIC.
This aligns with the findings in \cite{oldewage2023adversarial}.  
Since this method targets the black-box DNN detector through a white-box attack approach, the black-box DNN detector performs the worst after poisoning.
Meanwhile, the poisoning attack also significantly impacts the ResNet detector.
Fig. 10(b) illustrates the results of the 100-block data transmission, with SER averaged over five repetitions under varying SNR conditions. 
The poisoning attack effectively degrades the performance of all four types of deep receivers at different SNRs. 
Specifically, the attack degrades SER 
by 0.68 dB for the ResNet detector, 0.5 dB for the black-box DNN detector, 0.67 dB for DeepSIC, and 0.91 dB for Meta-DeepSIC at SNR = 10 dB.
As the SNR increases, the attack effect becomes more pronounced, 
with the SER deteriorations of the four receiver architectures reaching 0.68 dB, 0.43 dB, 1.40 dB, and 2.0 dB, respectively, at SNR = 15 dB.

\subsubsection{Non-linear Time-varying Synthetic Channels Results}

Fig. 11 evaluates the effectiveness of the proposed poisoning attack under a nonlinear time-varying synthetic channel,
using the same experimental setup as in Fig. 10.
The results indicate a stronger poisoning effect in the nonlinear channel compared to the linear one. 
Specifically, the attack degrades SER 
by 0.93 dB for the ResNet detector, 0.53 dB for the black-box DNN detector, 1.35 dB for DeepSIC, and 1.42 dB for Meta-DeepSIC at SNR = 10 dB. 
As the SNR increases, the attack effect becomes more pronounced, 
with the SER deteriorations of the four receiver architectures reaching 1.13 dB, 0.71 dB, 2.74 dB, and 3.09 dB, respectively, at SNR = 15 dB.

\begin{figure*}[!h]
  \centering
  \subfloat[\textrm{SER vs. block index, SNR = 14 dB}]
  {\includegraphics[width=3.1in]{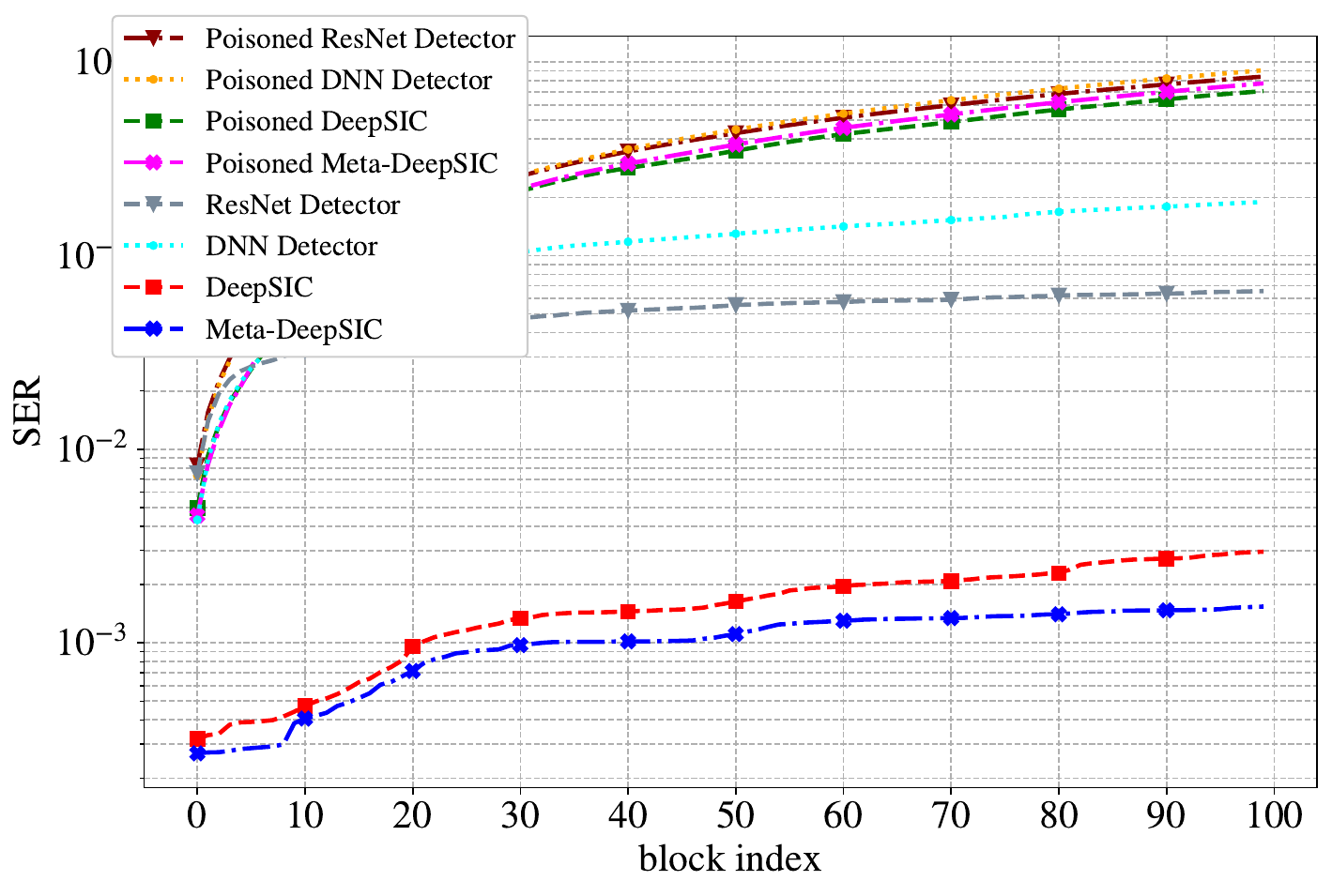}%
  \label{fig11_first_case}}
  \hfil
  \subfloat[\textrm{SER vs. SNR}]
  {\includegraphics[width=3.in]{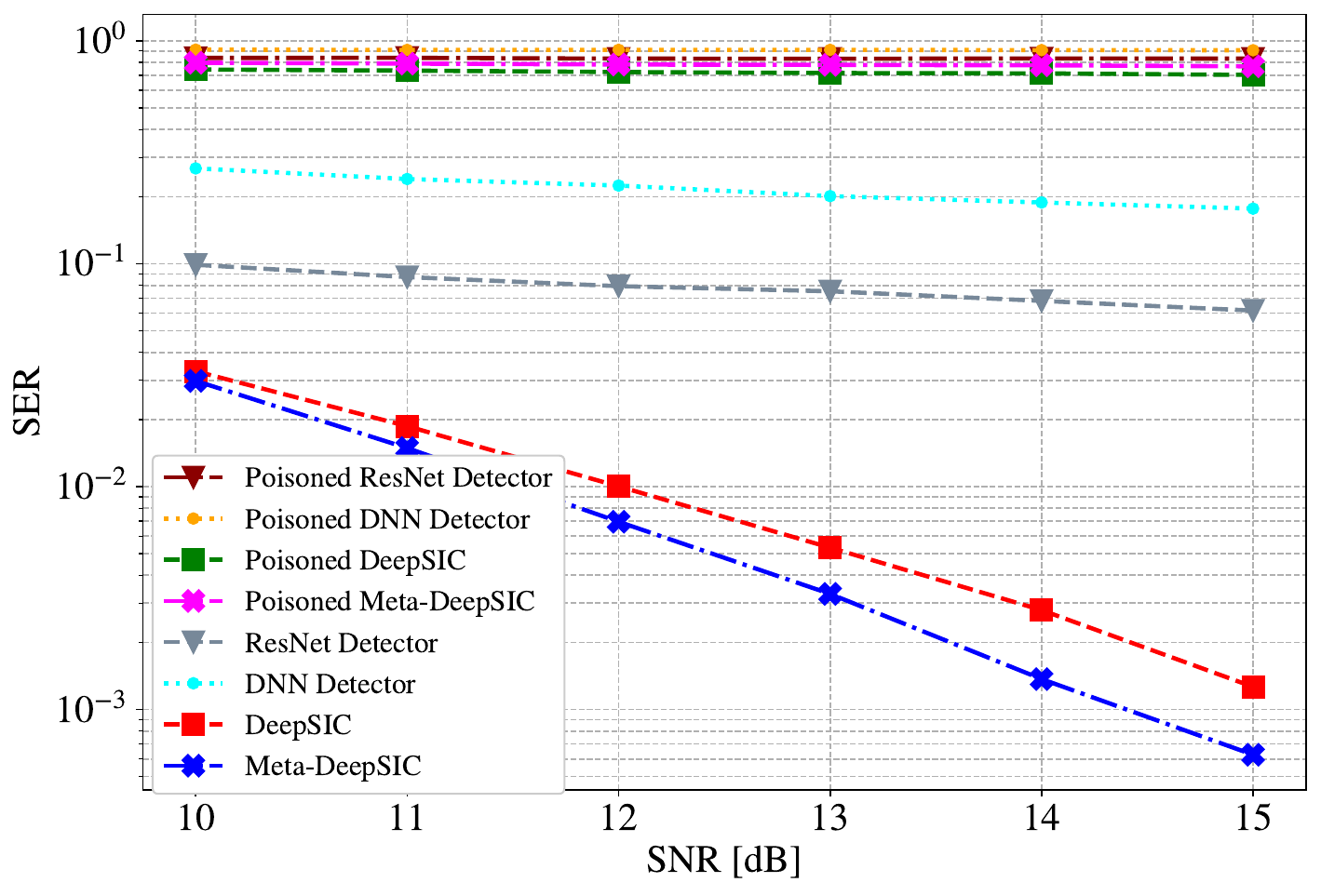}%
  \label{fig11_second_case}}
  \caption{Experimental results under nonlinear time-varying synthetic channels.}
  \label{fig11}
\end{figure*}

\subsubsection{Linear Static Synthetic Channels Results}

Fig. 12 evaluates the effectiveness of the proposed poisoning attack in a linear static synthetic channel, with the same setup as Fig. 10.
The performance of DeepSIC and Meta-DeepSIC is nearly identical in this case.
Furthermore, the lack of diverse data makes the data-driven detector unsuitable for adapting to the channel environment. 
For example, the black-box DNN detector struggles to adapt to the static channel, 
while the ResNet detector, which performed well in previous channels, performs the worst here, making the poisoning effect negligible.
In particular, at SNR = 10 dB, the poisoning attack degrades the SER by 0.08 dB for the black-box detector, 0.71 dB for DeepSIC, and 0.72 dB for Meta-DeepSIC. 
As the SNR increases, the impact intensifies, with SER degradations of 0.06 dB, 1.05 dB, and 1.07 dB at SNR = 15 dB, respectively.

\begin{figure*}[!h]
  \centering
  \subfloat[\textrm{SER vs. block index, SNR = 14 dB}]{\includegraphics[width=3.in]{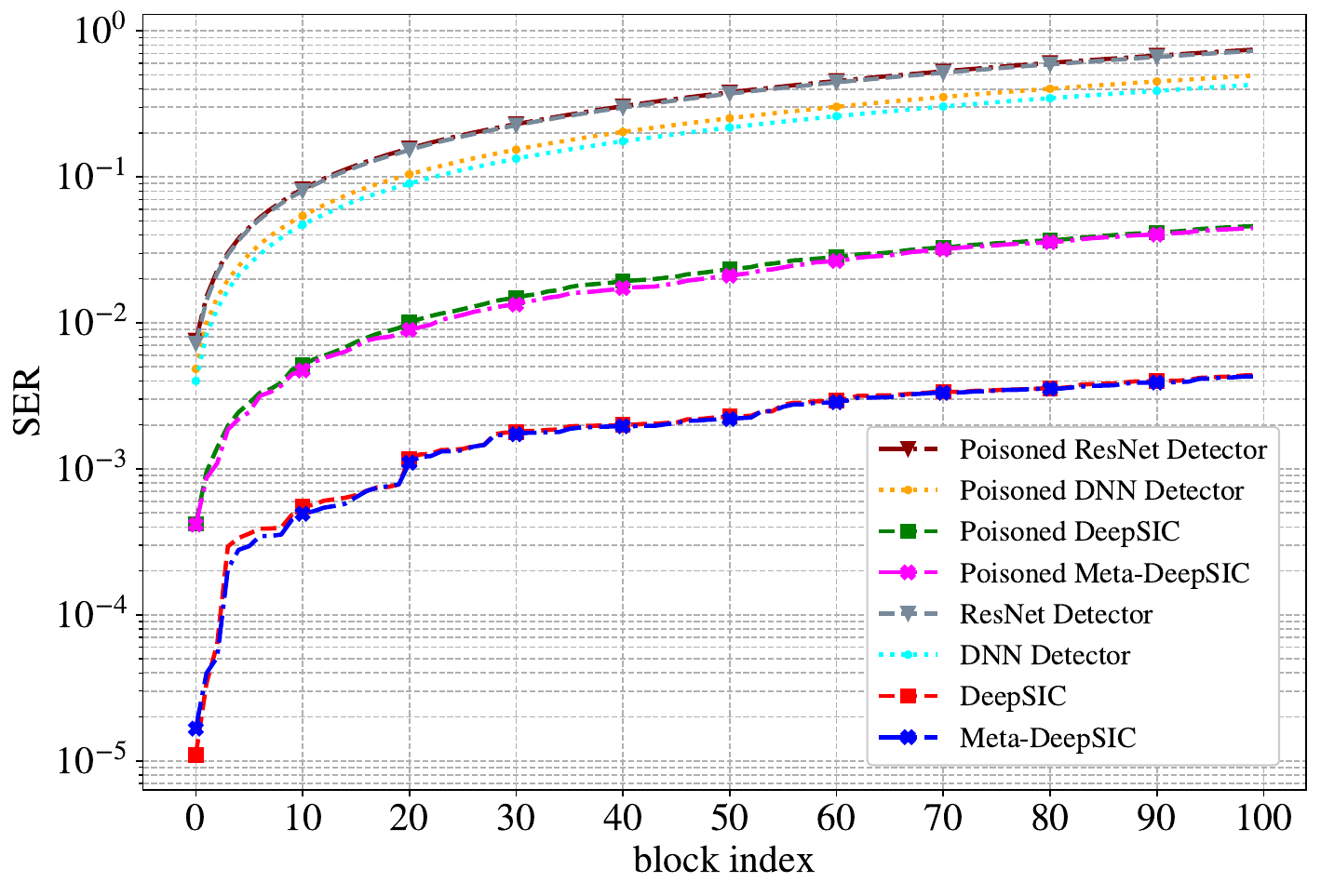}%
  \label{fig12_first_case}}
  \hfil
  \subfloat[\textrm{SER vs. SNR}]{\includegraphics[width=3.05in]{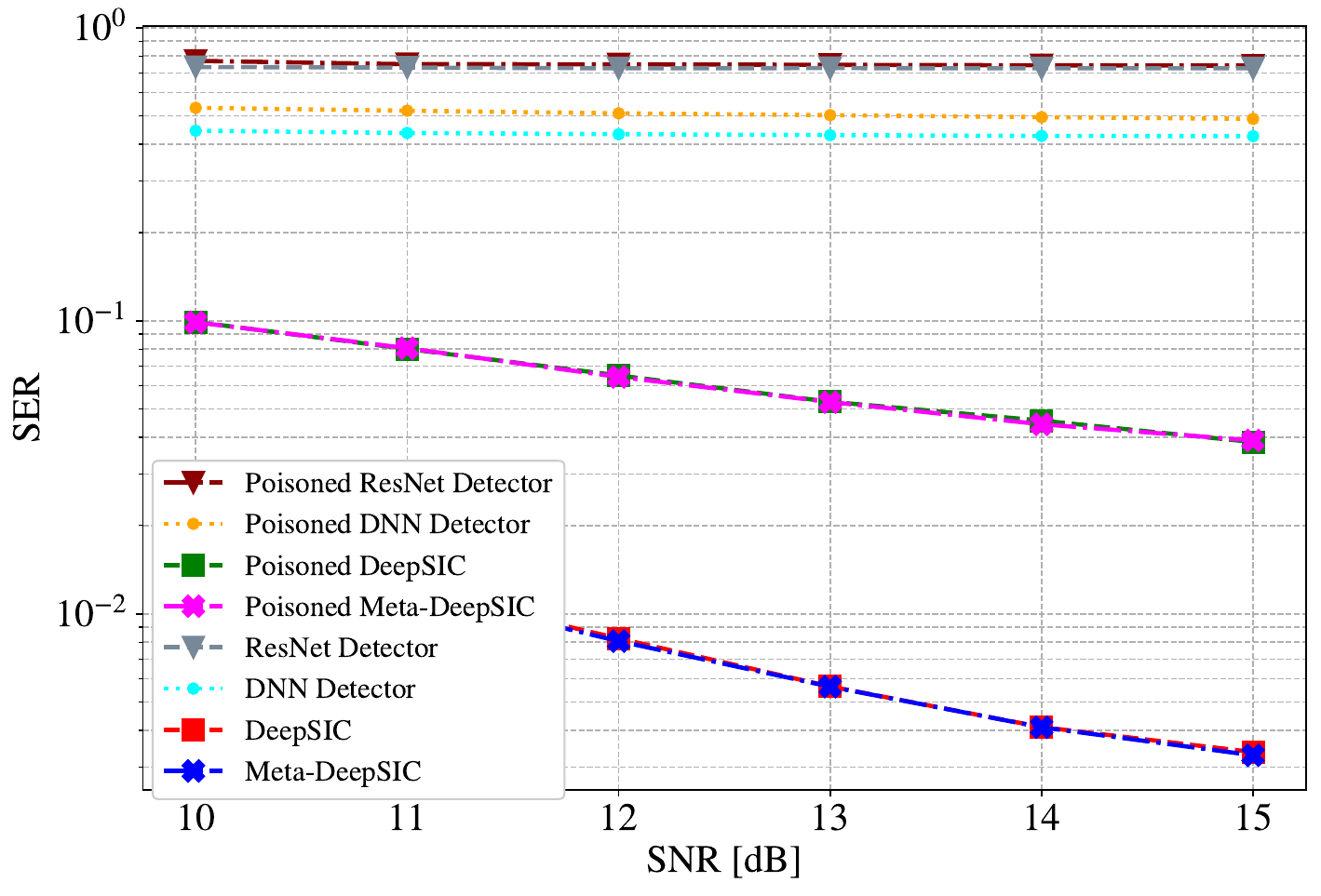}%
  \label{fig12_second_case}}
  \caption{Linear static synthetic channel experimental results.}
  \label{fig12}
\end{figure*}

\subsubsection{Time-varying COST 2100 Channels Results}

The effectiveness of the proposed poisoning attack is evaluated under the time-varying COST 2100 channel, as shown in Fig. 13. 
The experimental configuration is identical to that of the linear time-varying synthetic channel. 
Simultaneously, the surrogate model is trained on the joint channel data based on the time-varying linear channel model. 
As Fig. 13 illustrates, the poisoning attack proves effective across receivers with varying architectural and online training scenarios, 
while also adapting to different channel environments. 
At SNR = 10 dB, the attack degrades SER performance 
by 0.12 dB for the ResNet detector, 0.25 dB for the black-box DNN detector, 0.61 dB for DeepSIC, and 0.78 dB for Meta-DeepSIC. 
At SNR = 15 dB, the attack degrades SER performance 
by 0.04 dB for the ResNet detector, 0.23 dB for the black-box DNN detector, 2.11 dB for DeepSIC, and 1.36 dB for Meta-DeepSIC.

\begin{figure*}[!t]
  \centering
  \subfloat[\textrm{SER vs. block index, SNR = 14 dB}]{\includegraphics[width=3.in]{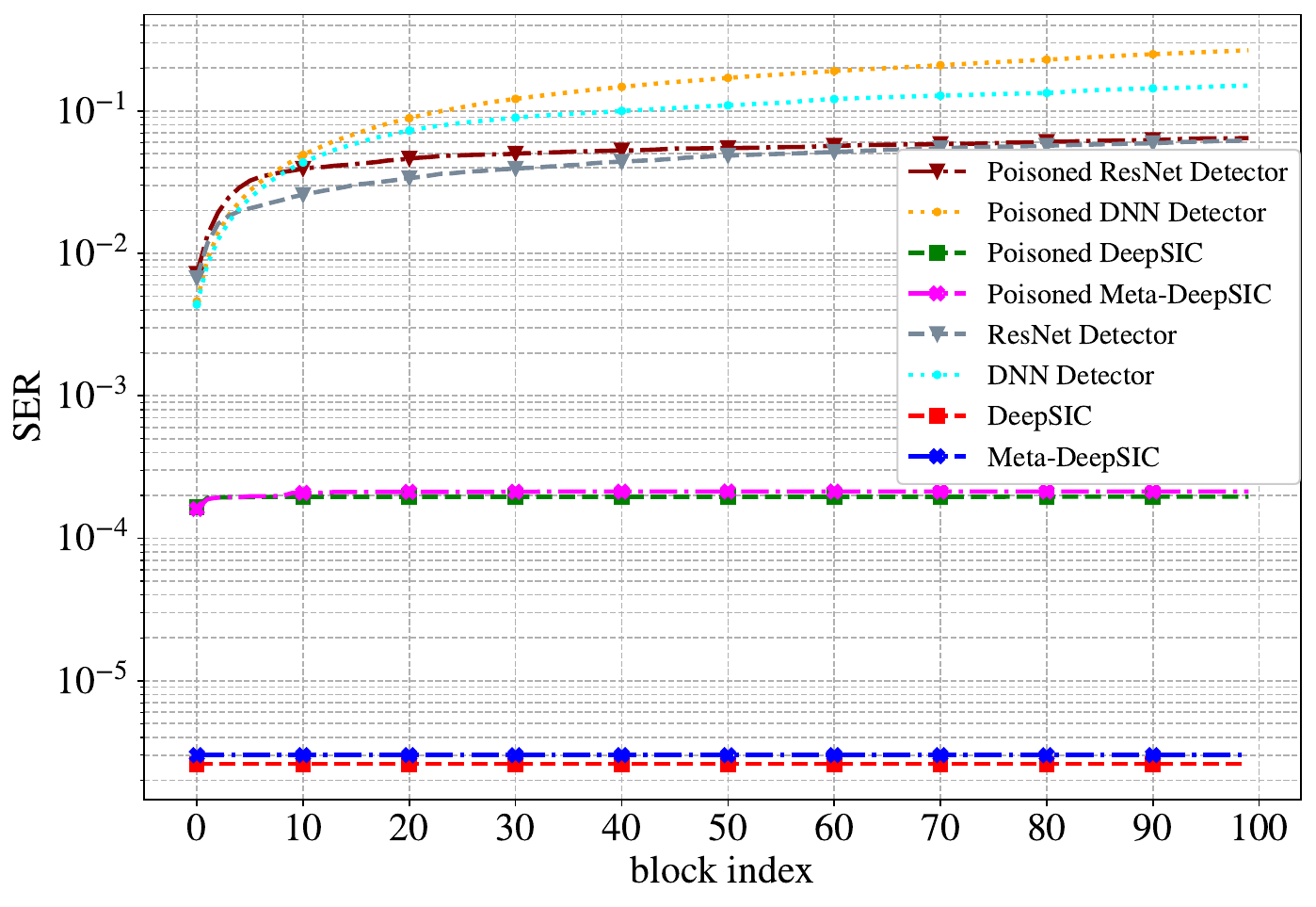}%
  \label{fig13_first_case}}
  \hfil
  \subfloat[\textrm{SER vs. SNR}]{\includegraphics[width=3.in]{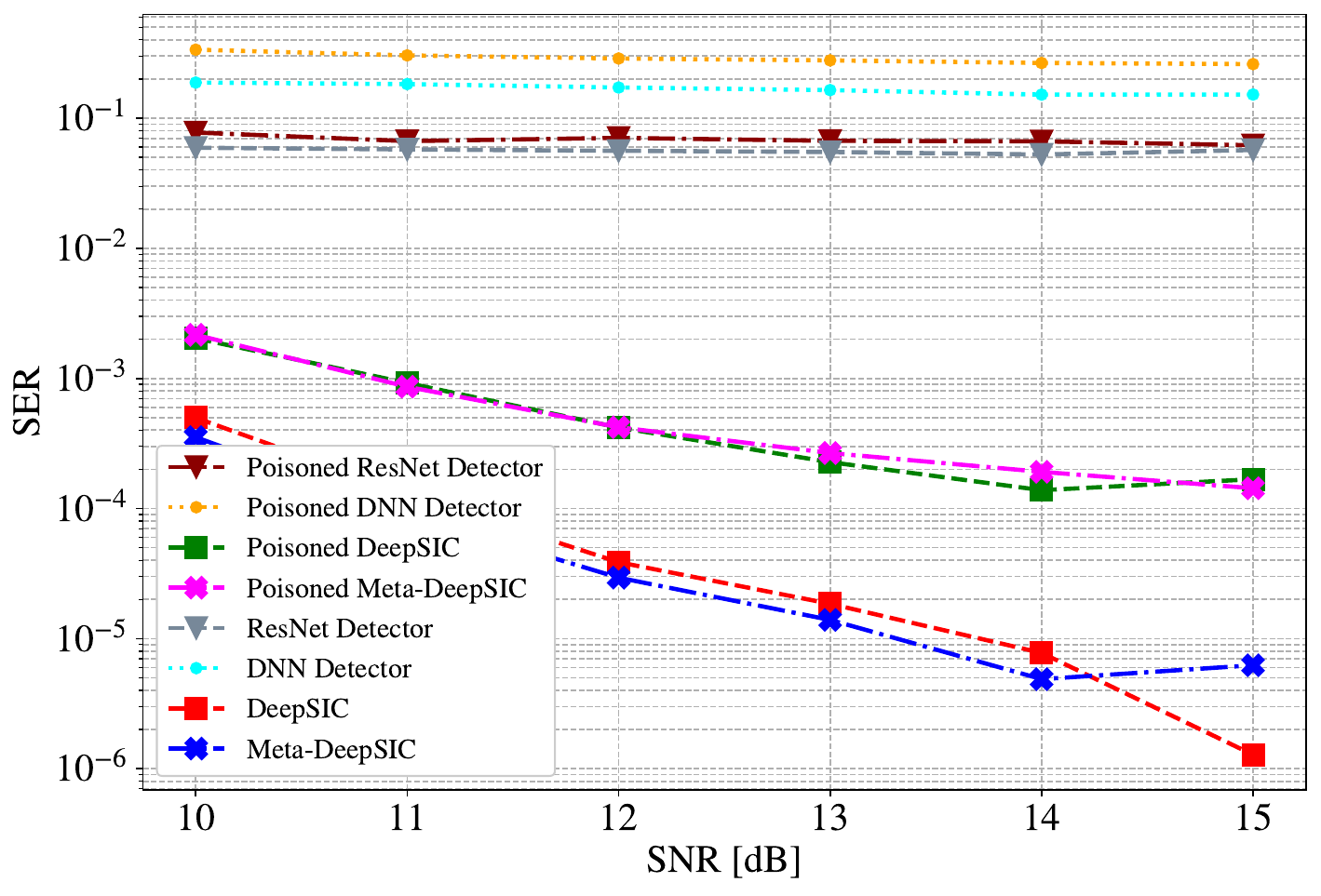}%
  \label{fig13_second_case}}
  \caption{Time-varying COST 2100 channel results.}
  \label{fig13}
\end{figure*}

\subsubsection{The Impact of Pilot Size on Poisoning}

\begin{figure}[!h] 
 \centering 
 \includegraphics[width=0.43\textwidth]{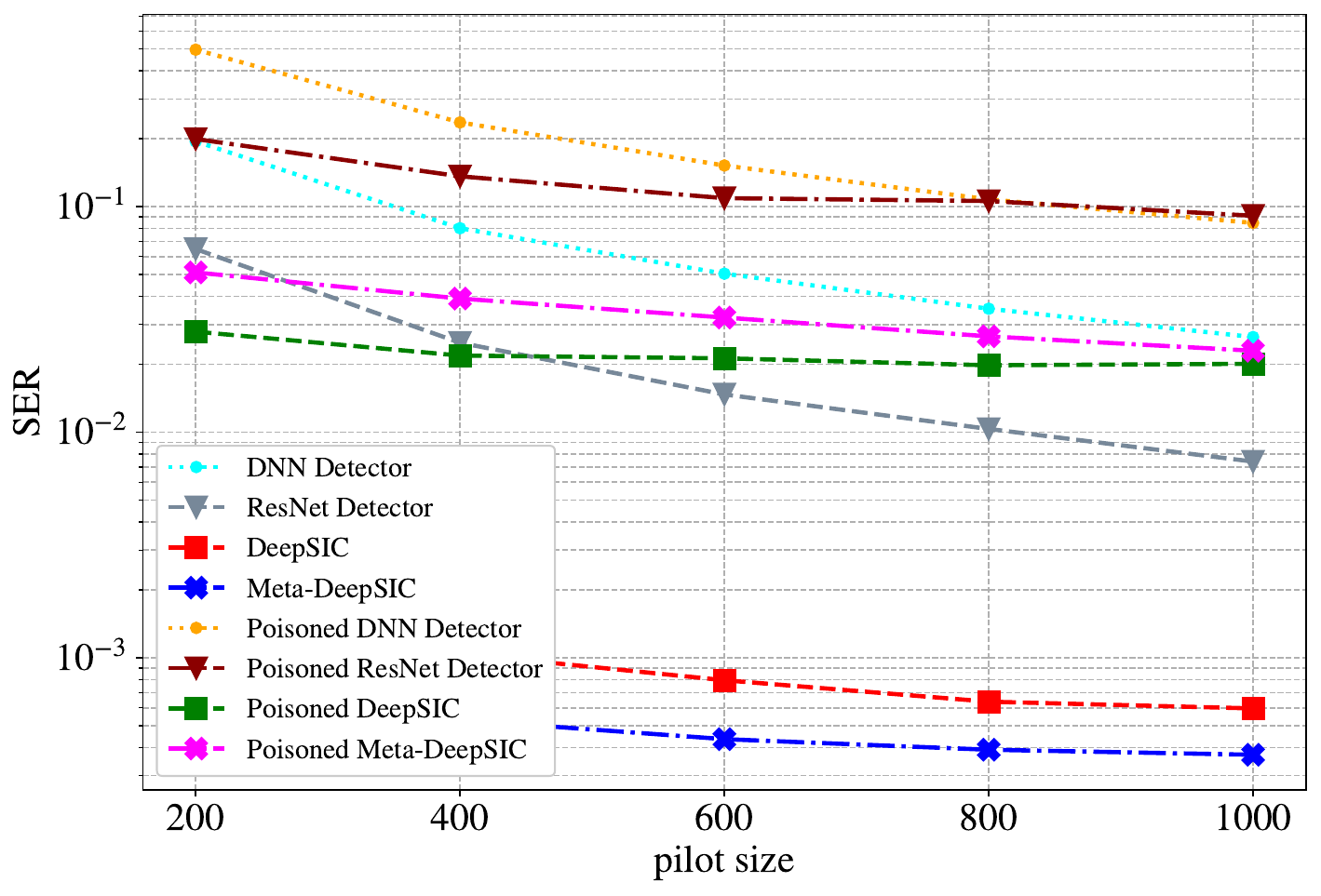} 
 \caption{SER vs. pilot size under Linear Time-varying Synthesis Channel (SNR = 14 dB)}
 \label{fig_14} 
\end{figure}

The online learning of deep receivers is influenced by the training data, i.e., the pilot size $L_{\text {pilot}}$. 
Larger pilot sizes generally enhance performance, leading to more stable learning and reduced overfitting.
To explore the impact of overfitting on the poisoning effect, 
Fig. 14 examines the influence of pilot size in a linear time-varying synthetic channel at SNR = 14 dB.
The results show that as $L_{\text {pilot}}$ increases, 
the effectiveness of poisoning attacks diminishes. 
Specifically, when comparing $L_{\text {pilot}}=200$ with $L_{\text {pilot}}=1000$, 
the poisoning effects on the ResNet detector, black-box DNN detector, DeepSIC, and Meta-DeepSIC are reduced by 0.34 dB, 0.74 dB, 0.14 dB, and 0.34 dB, respectively.
However, Larger pilot sizes cannot alleviate the poisoning effects on meta-learning methods, and the poisoned Meta-DeepSIC still performs worse than poisoned DeepSIC.

\subsection{Discussion of Results}
The experimental results demonstrate that the proposed poisoning attack is effective across four different channel environments. 
Nevertheless, the precise impact of these attacks varies depending on the specific channel environment. 
Firstly, as illustrated in Fig.11, the poisoning effect observed in the nonlinear time-varying synthetic channel is markedly higher than in the other environments. 
The performance of the deep receivers subjected to a poisoning attack is severely degraded and approaches failure in this channel environment. 
This suggests that the poisoning attack is capable of impeding the deep receiver's ability to adapt to the rapid changes in the channel's effects 
and to learn the nonlinear effects.

Secondly, the results for the linear time-varying synthetic channel and the COST 2100 channel support this conclusion from different perspectives.
In comparison to the synthetic channel, the tap coefficients of the COST 2100 channel demonstrate greater long-term variance, 
while exhibiting a relatively flat profile in the short term (e.g., between two blocks).
As illustrated in Fig. 13(a), the impact of poisoning attacks gradually reduces receiver performance in the COST 2100 channel, 
indicating a sustained but limited exacerbation. 
In contrast, in the linear time-varying synthetic channel undergoing significant changes over time, 
the poisoning attack has a more pronounced effect on the receiver's performance. 
This indicates that the attack has a impact on the deep receiver's capacity to track long-term channel alterations and 
a more pronounced disruptive effect on short-term rapid adaptation.
Notably, as shown in Fig. 14, a larger pilot size can mitigate the adverse effects of poisoning attacks, but this means more spectrum resources are consumed.


Finally, Meta-DeepSIC, which incorporates a meta-learning approach,
demonstrates optimal performance with limited pilot data, particularly in fast-varying channels.
Additionally, this learning capability is effective in nonlinear environments, 
but its increased sensitivity to poisoned samples leads to more significant performance deterioration compared to DeepSIC in fast-varying channels.
In slow-varying or static channels,  
the performance of both Meta-DeepSIC and DeepSIC, with or without poisoning attacks, is more stable than in fast-varying channels, as shown in Fig. 12 and Fig. 13.

In conclusion, the proposed attack primarily impedes the receiver's ability to learn rapid channel changes and non-linear effects 
in the short term, leading to performance degradation. 
Meanwhile, when deep receivers are combined with meta-learning, the poisoning effect is particularly pronounced (e.g., Meta-DeepSIC).
This conclusion highlights the security risks associated with designing wireless receivers using 
online and online meta-learning methods, particularly in environments characterised by rapid channel changes and non-linear effects, 
where the system is particularly susceptible to poisoning attacks.

\section{Conclusion}
This paper proposes a transfer-based adversarial poisoning attack method for online deep receivers without the knowledge of the target. 
The fundamental concept is to corrupt the online training and updating phases of deep receivers 
in such a way that the model becomes compromised after a designated period of training, 
resulting in a decline in performance. 
The poisoning attack framework and the generation of poisoning attack samples comprise two steps. 
Initially, the malicious user acquires the surrogate model through the joint learning method. 
Subsequently, the poisoning attack perturbations are generated based on the surrogate model to poisoning the pilot. 
Simulation experiments on the proposed poisoning attack method under varying channel models demonstrate that 
it disrupts the adaptation of dynamic channls and learning of nonlinear effects.
Meanwhile, the proposed attack can be effective against both model-based deep learning architectures and typical DNN-based receiver architectures.
Meta-DeepSIC demonstrates optimal performance in fast-varying channels. 
However, it is particularly susceptible to poisoning attack samples, resulting in a notable decline in performance. 
It is therefore recommended that future research should concentrate on the development of efficient, robust and secure deep receiver architectures 
that are capable of defending against potential attacks, 
such as poisoning purification before learning or reducing the impact after poisoning\cite{wang2024temporal},
with a view to furthering the application of deep learning in wireless transceiver design and deep receiver deployment.


\bibliographystyle{IEEEtran}
\bibliography{reference}

\newpage

\vspace{11pt}

\vspace{11pt}

\vfill

\end{document}